\documentclass[12pt,preprint]{aastex}    

\newcommand{\kms}{km s$^{-1}$}
\newcommand{\aforty}{$\alpha$.40}
\newcommand{\Jykms}{Jy$\:$km$\,$s$^{-1}$}
\newcommand{\htwo}{H$_2$}

\begin{document}

\title{A direct measurement of the baryonic mass function of galaxies \& implications for the galactic baryon fraction}

\author {Emmanouil Papastergis\altaffilmark{1}, Andrea Cattaneo\altaffilmark{2}, Shan Huang\altaffilmark{1}, Riccardo Giovanelli\altaffilmark{1} \& Martha P. Haynes\altaffilmark{1}}

\altaffiltext{1}{Center for Radiophysics and Space Research, Space Sciences Building,
Cornell University, Ithaca, NY 14853, USA. {\textit{e-mail:}} papastergis@astro.cornell.edu, shan@astro.cornell.edu, riccardo@astro.cornell.edu, haynes@astro.cornell.edu}

\altaffiltext{2}{Laboratoire d'Astrophysique de Marseille, UMR 6110 CNRS, Univ. d'Aix-Marseille, 38 rue F. Joliot-Curie, 13388 Marseille cedex 13, France  {\textit{e-mail:}} andrea.cattaneo@oamp.fr}

\begin{abstract}
We use both an HI-selected and an optically-selected galaxy sample to directly measure the abundance of galaxies as a function of their ``baryonic'' mass (stars + atomic gas). Stellar masses are calculated based on optical data from the Sloan Digital Sky Survey (SDSS) and atomic gas masses are calculated using atomic hydrogen (HI) emission line data from the Arecibo Legacy Fast ALFA (ALFALFA) survey. By using the technique of abundance matching, we combine the measured baryonic function (BMF) of galaxies with the dark matter halo mass function in a $\Lambda$CDM universe, in order to determine the galactic baryon fraction as a function of host halo mass. We find that the baryon fraction of low-mass halos is much smaller than the cosmic value, even when atomic gas is taken into account. We find that the galactic baryon deficit increases monotonically with decreasing halo mass, in contrast with previous studies which suggested an approximately constant baryon fraction at the low-mass end. We argue that the observed baryon fractions of low mass halos cannot be explained by reionization heating alone, and that additional feedback mechanisms (e.g. supernova blowout) must be invoked. However, the outflow rates needed to reproduce our result are not easily accommodated in the standard picture of galaxy formation in a $\Lambda$CDM universe.  
\end{abstract}

\maketitle

\section{Introduction}

It is by now well established that baryonic matter represents only about $1/6$ of the total matter density of the universe \citep[e.g.][]{Komatsu2011}, while the majority is in the form of non-baryonic dark matter (DM). Since galaxies form through the accretion of baryonic material onto dynamically dominant DM structures (halos), it would be reasonable to assume that the baryon mass fraction of present day galaxies approximately equals the cosmic value ($f_b = \Omega_b / \Omega_m \approx 0.16$). Despite this expectation, observations point to the fact that galaxies are not able to retain their cosmic ``fair share'' of baryons, and that the resulting baryon deficit depends strongly on the mass of their host halo.

The first line of evidence is provided by observational estimates of the abundance of galaxies as a function of their total stellar mass, a distribution referred to as the galactic stellar mass function (SMF). Thanks to the advent of wide area optical surveys with multiband photometric and spectroscopic information, such as the Two degree Field Galaxy Redshift Survey (2dFGRS) and the Sloan Digital Sky Survey (SDSS), the SMF has been measured over the mass range $M_\ast \approx 10^7 - 10^{12} \: M_\odot$, using statistical samples of tens of thousands of galaxies and a variety of stellar mass estimation techniques (\citealp{Cole2001}, \citealp{Bell2003}, \citealp{Panter2007}, \citealp{Baldry2008}, \citealp{LW2009}, \citealp{Yang2009}, \citealp{Baldry2012}, to name a few). The SMF displays an exponential cutoff at masses $M_\ast \gtrsim 10^{11} M_\odot$ and an approximate  power-law behavior at low masses ($dn \propto M_\ast^{-\alpha} \: dM_\ast $), with a ``shallow'' exponent of $\alpha \approx -1.3$. On the other hand, the halo mass function (HMF) predicted in the lambda cold dark matter ($\Lambda$CDM) model, follows a much ``steeper'' power-law  ($\alpha \approx -1.8$) over the mass range of interest. This observation alone excludes the possibility that the stellar mass of a galaxy is simply a fixed fraction of the host halo mass. 

One can furthermore statistically derive an average relation between the stellar mass of a galaxy ($M_\ast$) and the mass of its host halo ($M_h$), through the technique of abundance matching (see \S \ref{sec:am} for details). $M_\ast$ - $M_h$ relations based on abundance matching \citep[e.g.][]{Guo2010, Moster2010, Behroozi2010, Leauthaud2012} have shown that the ``stellar conversion efficiency'', $\eta_\ast = (M_\ast/M_h) \, /  \, f_b $, never exceeds 25 - 30\%. Furthermore, $\eta_\ast$ peaks for Milky Way-sized galaxies ($M_h \approx 10^{12} M_\odot$), and declines rapidly on either side of the peak (e.g. Figure 2 in \citealp{Guo2010}).

The second line of evidence comes from direct halo mass measurements, obtained through weak lensing or kinematics studies \citep[e.g.][]{Dutton2010, Reyes2012}. For example, \citet{Reyes2012} used stacked weak lensing measurements to estimate the average host halo mass of galaxies in different stellar mass bins, and found that $\eta_\ast$ never exceeds $\approx 30$\%. Direct halo mass measurements can circumvent a number of assumptions inherent in the application of abundance matching, but such techniques can presently only be applied to a restricted range of stellar mass ($M_\ast \approx 10^9 - 10^{11} \: M_\odot$), and are affected by their own set of systematics. 

Stellar mass is not always the dominant baryonic component in a galaxy. In fact, the HI--to--stellar mass ratio (``HI fraction''; $f_{HI} = M_{HI} / M_\ast$) tends to increase with decreasing stellar mass, and HI often dominates the baryonic content of low-mass galaxies. The transition from stellar-mass--dominated to HI--dominated systems takes place at $M_\ast \approx 10^{10} M_\odot$ for HI-selected samples \citep[e.g.][see also Fig. \ref{fig:gsmr} in this work]{Huang2012b}, or at $M_\ast \lesssim 10^{9.5} \: M_\odot$ for optically-selected samples \citep[e.g.][]{Catinella2010}. As a result, it is presently not clear what is the behavior of the ``baryon retention fraction'' $\eta_b = (M_b/M_h) \, / \, f_b$ in low-mass galaxies, when both stars and cold gas are taken into account. In particular it is not well understood whether the very low average value of $\eta_\ast$ inferred for low-mass halos is a result of poor retention of baryonic material, of the low efficiency of gas-to-stars conversion, or of a combination of both. 

For example, \citet{Baldry2008} argue that the increasing gas fraction in low-mass galaxies should approximately offset the decreasing stellar-to-halo mass ratio, and result in a roughly constant $\eta_b \approx 10$\%. This conclusion was based on an indirect estimate of the cold gas content of galaxies, based on the average $f_{HI} - M_\ast$ relation observed in a set of samples in the literature. An early work by \citet{SP1999}, based on the same indirect method, also reached a qualitatively similar conclusion. \citet{Evoli2011} found an approximately constant $\eta_b$ at the low-mass end using a different indirect method, which involves the comparison of the stellar and HI mass distributions of two different galaxy samples. These results would imply that low-mass galaxies are relatively efficient at retaining baryonic mass, but very inefficient in converting their  gas into stars. This conclusion, however, would require a ``steep'' HI mass function (HIMF) in the local universe, in contrast to what is measured \citep{Zwaan2005, Martin2010}. Moreover, the recent work of \citet{Rodriguez2011}, also based on using the average $f_{HI} - M_\ast$ relations for blue and red galaxies separately, found no signs for a flat $\eta_b$ at low masses.    
  
In this article we \textit{directly} measure the abundance of galaxies as a function of their ``baryonic mass'' (throughout this article the term baryonic refers to the combined stellar and atomic gas components of galaxies, and baryonic mass is calculated as $M_b = M_\ast + 1.4 \, M_{HI}$, where the 1.4 factor accounts for the presence of helium). We use optical data from the seventh data release of the SDSS (SDSS DR7) to estimate stellar masses, and HI-line flux measurements from the Arecibo Legacy Fast ALFA\footnotemark{} (ALFALFA) survey to measure atomic gas masses. The resulting distribution, referred to hereafter as the baryon mass function (BMF) of galaxies, can be used in abundance matching to derive a robust $\eta_b$ - $M_h$ relation. In order to investigate sample selection effects, we employ both an HI-selected and an optically-selected sample drawn from the \textit{same} volume to derive the mass distributions for the stellar, atomic hydrogen and baryonic components.

\footnotetext{The Arecibo L-band Feed Array (ALFA) is a 7-feed receiver operating in the L-band ($\approx 1420$ MHz), installed at the Arecibo Observatory.}

The paper is organized as follows: in section \ref{sec:data}, we introduce the datasets used to measure the stellar, HI and baryon mass distributions. We describe the methodology used to measure atomic hydrogen masses and we estimate stellar masses for our galaxy samples. In section \ref{sec:bmf}, we present our measurements of the SMF, HIMF \& BMF from both the HI-selected and the optically-selected samples, and compare them against one another as well as against other published results. In section \ref{sec:biases} we consider the impact of possible systematics on our measurements, such as stellar mass estimation method, distance uncertainties and the exclusion of some baryonic components (e.g. molecular gas) in the calculation of the BMF. In section \ref{sec:bar_frac}, we present the $\eta_\ast$ - $M_h$ and $\eta_b$ - $M_h$ relation in a $\Lambda$CDM universe. In section \ref{sec:conclusion}, we discuss the implications of the result and summarize our main conclusions. Throughout this paper, we use a Hubble constant of $H_0 = 70 \: h_{70} \;\; \mathrm{ km}\, \mathrm{s}^{-1}  \mathrm{Mpc}^{-1}$.

\section{Datasets \& derived quantities   \label{sec:data}}

\subsection{HI-selected sample  \label{sec:hi_selected}}

We select galaxies from the current data release of the ALFALFA survey, which covers 40\% of the planned final survey area (``\aforty'' catalog; \citealp{Haynes2011}). We restrict ourselves to two rectangular areas of the ``spring'' coverage of \aforty \ ($07^h45^m < RA <16^h30^m$, $4^\circ < Dec < 16^\circ$ \& $24^\circ < Dec < 28^\circ$), which encompass the Virgo cluster as well as the supergalactic plane at low velocities. We restrict ourselves to galaxies with $v_{CMB} < $ 15000 \kms \ ($z<0.05$), in order to avoid the strong radio frequency interference (RFI) present at frequencies that correspond to $v_\odot \gtrsim 15000$ \kms. We discard the nearest extragalactic sources with $D<10$ Mpc, because they can carry extreme fractional uncertainties on their distances (see \S\ref{sec:derived} for details on the distance assignment method). We furthermore select only HI sources designated as ``Code 1'' in \aforty, i.e. extragalactic sources detected at high significance ($S/N_{HI} > 6.5$). In addition, we exclude sources with integrated fluxes below the 50\% completeness limit of the ALFALFA survey (see \citealp{Haynes2011}, Section 6 for the derivation of the ALFALFA completeness limits). The above requirements are satisfied by 7618 galaxies.

We remove from our sample 204 \aforty \ sources which are not crossmatched with an optical source in SDSS, as well as 208 additional sources which have been flagged as having problematic SDSS photometry (crossmatch code ``P'' in \aforty, for details see Section 4 in \citealp{Haynes2011}). This quality cut on the SDSS photometry introduces some bias against faint, low surface brightness galaxies of irregular morphology; such sources are often ``shredded'' (i.e. assigned multiple photometric objects) by the SDSS magnitude extraction process, and are usually assigned a ``P'' (``photometry suspect'') crossmatch code in \aforty. Lastly, 11 additional objects were discarded, in cases where the stellar mass computation method described in \S \ref{sec:derived} failed to produce physically plausible results.

\begin{figure}[htbp]
\includegraphics[scale=0.75]{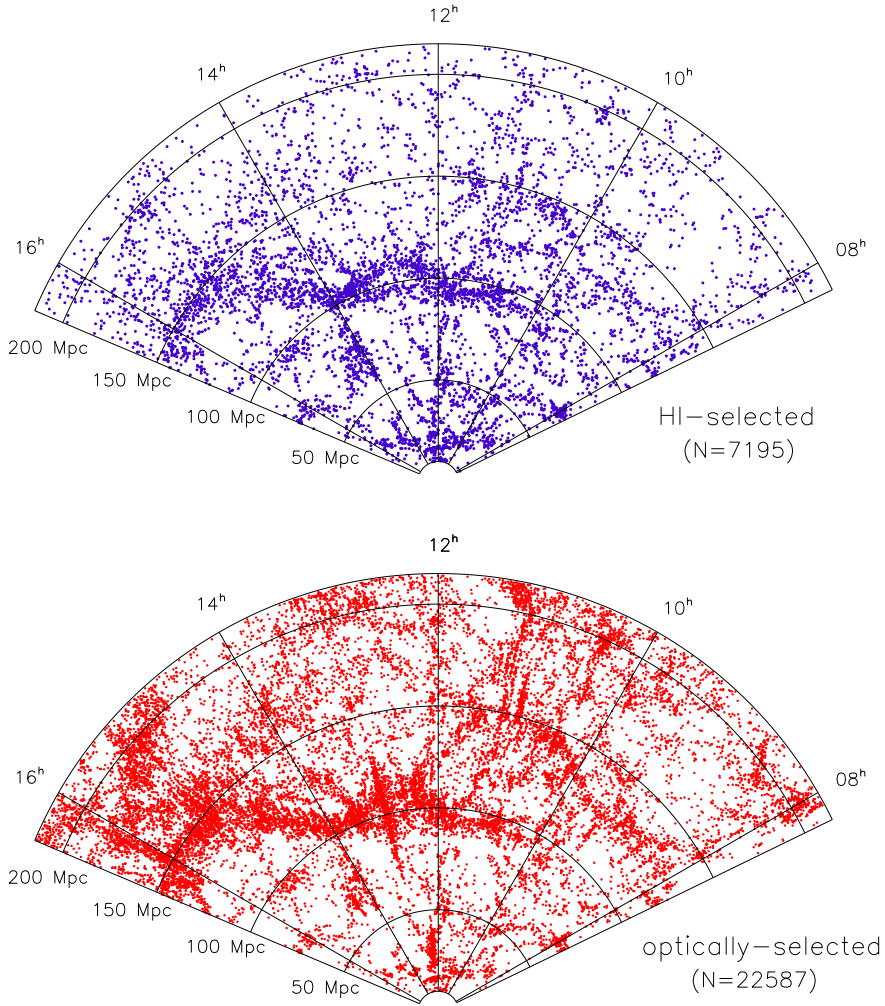}
\caption{Spatial distribution of the 7195 HI-selected galaxies (\textit{upper panel}) and 22587 optically-selected galaxies (\textit{lower panel}), drawn from the same volume. The galaxy stellar mass function (SMF), HI mass function (HIMF) and baryonic mass function (BMF) are computed separately for the two samples, in order to assess the impact of sample selection on the derived distributions.}
\label{fig:coneplot}
\end{figure}

Our final sample thus consists of 7195 extragalactic objects, detected over $\approx$2000 deg$^2$ of high Galactic latitude sky and out to $D\approx 214$ Mpc. The upper panel of Figure \ref{fig:coneplot} displays the spatial distribution of our HI-selected galaxies, and puts in evidence the complex large scale structure in the survey volume. Note that all objects in our HI-selected sample have 21cm redshifts \footnotemark{} and line fluxes as well as multi-band optical photometry, and hence estimates of both their stellar and atomic hydrogen masses.

\footnotetext{Of the 7195 galaxies in the HI-selected sample, 1333 are not in the SDSS DR7 spectroscopic database and thus lack SDSS optical redshifts.}

\subsection{Optically-selected sample  \label{sec:opt_selected}}

We draw an optically-selected sample from the SDSS DR7 \citep{Abazajian2009} spectroscopic database, in the same volume used to define our HI-selected sample. Specifically, we select galaxies that lie within the same sky area ($07^h45^m < RA <16^h30^m$, $4^\circ < Dec < 16^\circ$ \& $24^\circ < Dec < 28^\circ$) and satisfy the same velocity and distance restrictions ($v_{CMB}<$ 15000 \kms, $D > 10$ Mpc). CMB velocities for our optically-selected galaxies are calculated based on their SDSS spectroscopic redshifts ($z_{SDSS}$).  We restrict ourselves to objects spectroscopically classified as galaxies in SDSS (\texttt{specClass = 2}) that also have an apparent Petrosian magnitude brighter than 17.5 in the $r$-band ($r_{petro} < 17.5$). This initial cut results in 22707 galaxies. Due to their large number, it is not practical to inspect all galaxies individually for the quality of their SDSS photometry/spectroscopy. As a result, we expect a fraction of our sources to have issues with their SDSS photometry, most often related to ``shredding'' (i.e. assignment of multiple photometric objects to a single galaxy). This issue affects mostly extended sources with structure in their light distribution, such as low surface brightness (LSB) galaxies with irregular morphology. In such cases, the SDSS magnitude will underestimate the true flux of the galaxy, which in turn will result in an underestimate of its stellar mass. A second issue related to shredding, is that bright star forming knots in the disks of nearby spiral galaxies can sometimes be cataloged as separate spectroscopic objects, and hence interpreted as low-mass satellites of the main spiral. We find that applying a color cut on our sample,  $(i-z)_{model} > -0.25$, removes a fair fraction of these unwanted cases. On the other hand, cuts based on the quality of the SDSS spectrum (such as cuts on \texttt{zconf}, \texttt{zstatus} or \texttt{zwarning}) are ineffective, since they exclude mostly legitimate faint or LSB dwarf galaxies with noisy spectra. Lastly, we exclude objects for which the stellar mass computation described in \S \ref{sec:derived} failed to produce physically plausible results.   

Our final optically-selected sample consists of 22587 galaxies, occupying the same volume as our HI-selected sample. We crossmatch the optical sample with the full \aforty \ catalog (including Code 1 \& 2 sources), and find 7551 HI source counterparts. The crossmatch rate is thus approximately $1/3$, as reported in \citet{Haynes2011}. Note that ALFALFA non-detected galaxies are not necessarily HI-poor objects; due to the low emissivity of atomic hydrogen in the 21cm line, even moderately gas-rich galaxies in the outer portion of the survey volume can be missed by ALFALFA. This point is illustrated by the lower panel of Figure \ref{fig:coneplot}, which compares the spatial distribution of galaxies in the HI-selected and optically-selected samples. Note that all optically-selected galaxies have multiband optical photometry as well as optical redshifts, and hence an estimate of their stellar mass. However, only galaxies crossmatched with an \aforty \ source have a 21cm flux measurement, and hence a value for their atomic hydrogen mass.

\subsection{Derived quantities \label{sec:derived}}

\begin{figure}[htbp]
\includegraphics[scale=0.75]{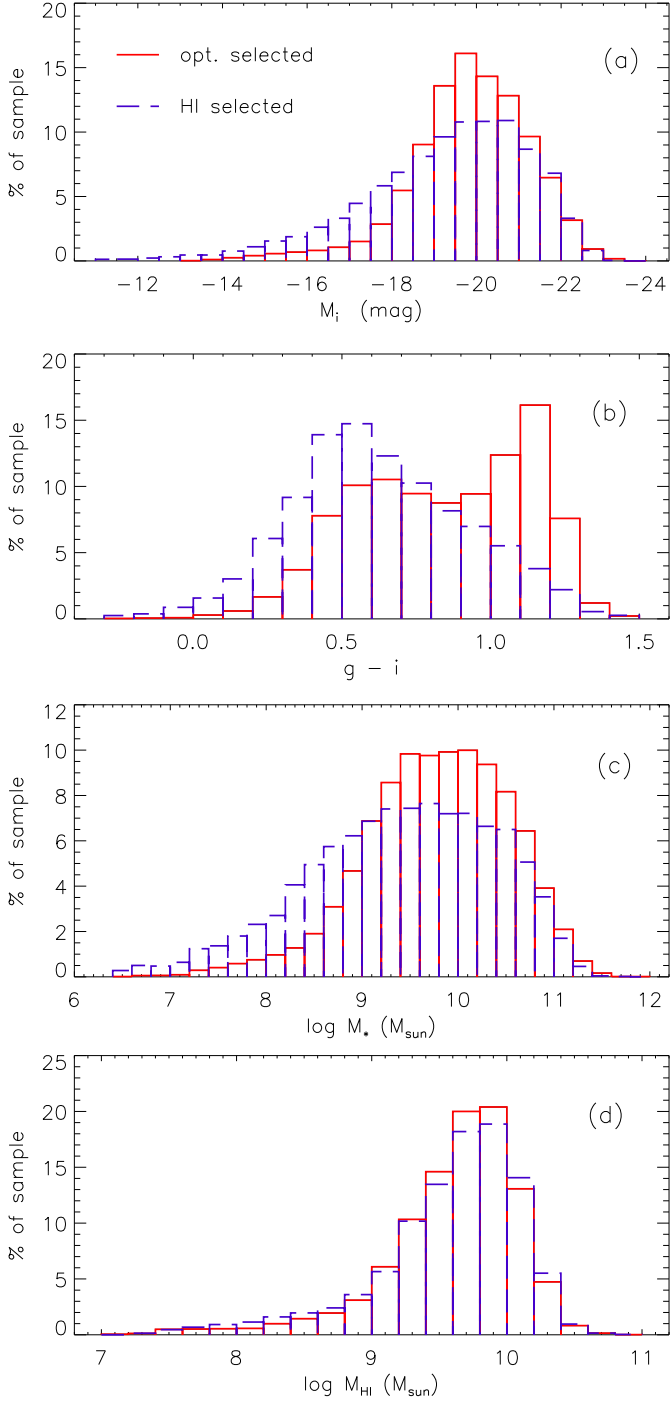}
\caption{ Histograms of $i$-band absolute magnitude (\textit{panel a}), $g-i$ color (\textit{panel b}), stellar mass (\textit{panel c}) and HI mass (\textit{panel d}), for the optically-selected (\textit{red solid line}) and HI-selected (\textit{blue dashed line}) samples. As evident in panel \textit{b}, the HI-selected sample is strongly biased against the red galaxy population and as a result it is skewed towards lower luminosity and stellar mass systems. Conversely, the fractional contribution of bright and massive galaxies ($M_i \lesssim -19, \; M_\ast \gtrsim 10^9 M_\odot$) is larger for the optically-selected sample (panels \textit{a} and \textit{c}). } 
\label{fig:histogram}
\end{figure}

We calculate HI masses from the measured 21cm integrated flux reported in \aforty. Assuming optically thin emission

\begin{equation}
M_{HI} = 2.356\,10^5\; S_{int}\; D^2 \; ,
\end{equation}

\noindent
where $M_{HI}$ is the HI mass in units of the solar mass ($M_\odot$), $S_{int}$ is the integrated flux in \Jykms \ and $D$ is the distance in Mpc. Distances in this article are calculated according to the method used in \aforty \ \citep{Haynes2011}: nearby galaxies ($v_{CMB} < 6000$ \kms) are assigned distances through the use of a peculiar velocity flow model developed by \citet{Masters2005}, while for more distant galaxies simple Hubble distances are used ($D = v_{CMB}/H_0$, with $H_0 = 70$ km$\,$s$^{-1}$Mpc$^{-1}$). Moreover, group and cluster membership information is taken into account when available, as well as primary distance measurements published in the literature. We would like to point out that most of the galaxies in our optically-selected sample are not included in \aforty, and hence lack the systematic group assignments and primary distance information contained in the catalog. Nevertheless, optically-selected galaxies that lie within the sky area and redshift range of the Virgo cluster are placed collectively at the Virgo distance ($D = 16.5$ Mpc), in order to minimize the effects of  peculiar motions on the inferred distances of galaxies in the region. We would also like to note that the distance assignment method can have a large impact on the determination of mass functions, especially at the low-mass end. We illustrate this issue in \S\ref{sec:dist_unc}, where we consider the effect on the HIMF of using uniformly Hubble distances for all galaxies.

We compute stellar masses for our galaxies based on fitting all 5 SDSS photometric bands ($u,g,r,i,z$), with model spectral energy distributions (SEDs). The full details of the method can be found in \citet{Huang2012a}, but here we summarize the main points: a library of model SEDs are generated, using the \citet{BC2003} stellar population synthesis code and assuming a \citet{Chabrier2003} stellar initial mass function (IMF). Models with an extensive range of internal extinction, metallicity and star formation histories are considered. In particular, star formation history templates include both an exponentially declining component as well as random starburst episodes. The final physical properties (e.g. stellar mass, star formation rate, internal extinction etc.) are computed as the average of all model values, where each model is weighted according to its fit likelihood. In addition to mean values, ``$1\sigma$'' uncertainties of the physical properties can also be derived, as one quarter of the 2.5-97.5 percentile range of model values. The median $1\sigma$ uncertainty in $\log M_\ast$ is 0.086 dex, or about 22\% (excluding uncertainties on the distance). It is important to note that stellar mass estimates of the same galaxy obtained with different methods can have systematic offsets of up to factors of a few. In  \S\ref{sec:stellar_est} we address issues related to stellar mass estimation, and consider alternative methods for calculating stellar masses \citep{Bell2003,Taylor2011}. 

Figure \ref{fig:histogram} compares the distributions of $i$-band absolute magnitude ($M_i$) and $g-i$ color (both corrected for Milky Way extinction), stellar mass ($M_\ast$) and HI mass ($M_{HI}$) for the HI-selected and optically-selected samples. The most notable difference is in the $g-i$ color distribution, where the HI-selected sample shows a strong bias against the red galaxy population (see also \citealp{Huang2012b}). As a result, the optically-selected sample contains a larger proportion of high luminosity and stellar mass systems compared to the HI-selected sample. By contrast, the $M_{HI}$ distribution of the two samples is very similar, but remember that only those optically-selected galaxies that are detected in ALFALFA are included in the histogram.

\section{The baryonic mass function   \label{sec:bmf}}

\subsection{Method}
\label{sec:method}

Stellar masses for all galaxies, and HI masses for all ALFALFA-detected galaxies are calculated as described in \S \ref{sec:derived}. For the $\approx$15000 galaxies in the optically-selected sample that lack an ALFALFA detection, we assign a lower and an upper limit on their atomic hydrogen content ($M_{HI}^{min}, M_{HI}^{max}$). The lower limit is simply $M_{HI}^{min} = 0$, which corresponds to an HI-devoid galaxy. The upper limit is calculated by assuming that the HI flux of the non-detected galaxy lies just below the ALFALFA ``detection limit'', as defined by the 25\% completeness limit of the \aforty \ catalog when both Code 1 \& 2 sources are considered. More precisely

\begin{equation}
\log M_{HI}^{max} = 5.372 + \log{S_{int}^{25\% \, lim}} + 2 \log D \; ,
\end{equation} 

\noindent
where $D$ is the galaxy distance in Mpc determined by its SDSS optical redshift, and $S_{int}^{25\% \, lim}$ is the flux level at which the completeness of the \aforty \ catalog falls to 25\%, in Jy \kms. According to Eqns. 6 \& 7 of \citet{Haynes2011}

\begin{eqnarray}
\log{S_{int}^{25\% \, lim}} = 
\left\{
\begin{array}{lr}
0.5 \log{W_{50}} - 1.312 & \log{W_{50}} \, , \leqslant 2.5 \\
 \; \log{W_{50} - 2.562} & \log{W_{50}} \, , > 2.5 \\

\end{array}
\right.
\end{eqnarray}

\noindent
where $W_{50}$ is the HI-line profile width in \kms, measured at the 50\% flux level of the profile peak. Since ALFALFA non-detected galaxies lack a measurement of $W_{50}$, we assign a value based on the average $M_\ast$-- $v_{rot}$ relation (i.e. the stellar mass Tully-Fisher relation) of \aforty \  galaxies. We then project the $v_{rot}$ value on the line-of-sight according to the SDSS $r$-band axial ratio, and assuming an intrinsic axial ratio of $q_0 = 0.13$ for all galaxies.   

Baryonic masses (i.e. stellar mass + atomic gas mass) for all galaxies are calculated as $M_b = M_\ast + 1.4 \, M_{HI}$, where the 1.4 factor is used to account for the cosmic abundance of helium. Note that ALFALFA non-detected galaxies have two assigned values for their HI mass, and consequently two values for their baryonic mass, $M_b^{min} = M_\ast$ and $M_b^{max} = M_\ast + 1.4 \, M_{HI}^{max}$.
 
We calculate cumulative mass functions in logarithmic mass bins for all three components (i.e. stellar mass, atomic hydrogen mass, baryonic mass), separately for the HI-selected and optically-selected samples. Since neither sample is volume-limited, mass functions have to take into account the sample selection criteria as well as the large-scale structure in the survey volume. HI selection is based on a combination of galactic HI integrated flux, $S_{int}$, and profile width, $W_{50}$ (see \S \ref{sec:hi_selected} \& discussion in Section 6 of \citealp{Haynes2011}); as a result, galaxies of different HI masses and linewidths are detected out to different distances. Similarly, our optically-selected sample is a flux-limited sample, which results in galaxies with different $r$-band absolute magnitudes being detected in different volumes. As a result, mass distributions are calculated by summing up the number of detections in a given mass bin (see Fig. \ref{fig:histogram}), with each detection weighted by an appropriate volume factor. Individual weighting factors are calculated via the ``$1/V_{eff}$'' method, as implemented in \citealp{Zwaan2005}. This is a non-parametric, maximum-likelihood method, which reduces to the standard $1/V_{max}$ method \citep{Schmidt1968} when applied to a spatially homogeneous galactic sample. The advantage of $1/V_{eff}$ consists in the fact that it is insensitive to local density fluctuations, and hence mostly immune to structure-induced bias. We refer the interested reader to the literature for the definition, implementation and details of the method. The method definition and basic setup is described in \citet{Efstathiou1988}; details of the implementation of the method for HI data can be found in \citet{Zwaan2003, Zwaan2005}; the specifics of the application of the method on ALFALFA data can be found in \citet{Martin2010}, while a shorter qualitative description can be found in \citet{Papastergis2011}.

Lastly, a fraction of galaxies that satisfy all criteria for SDSS spectroscopic followup cannot be observed for technical reasons (mostly fiber collisions), and are therefore not included in the SDSS spectroscopic database. We therefore correct the normalization of all optically-selected distributions by $1/<f_{spec}>$, using the average spectroscopic completeness value reported in \citet{LW2009}, $<f_{spec}> = 0.92$. Similarly, a fraction of the ALFALFA volume is ``lost'' due to RFI contamination of certain frequency bands in the ALFALFA passband. We correct the normalization of all HI-selected distributions by $1/ (1-f_{RFI})$, where $f_{RFI} = 0.03$. We would also like to note that, due to the $4^\prime$ beam size of the ALFA receiver, a number of HI sources are expected to be blended. We do not attempt to correct for blending but, given that HI-selected galaxies are a weakly clustered population \citep[e.g.][]{Martin2012}, we anticipate the effect on the HI-selected distributions to be small.

\subsection{Results   \label{sec:results}}


\begin{figure}[htbp]
\includegraphics[scale=0.75]{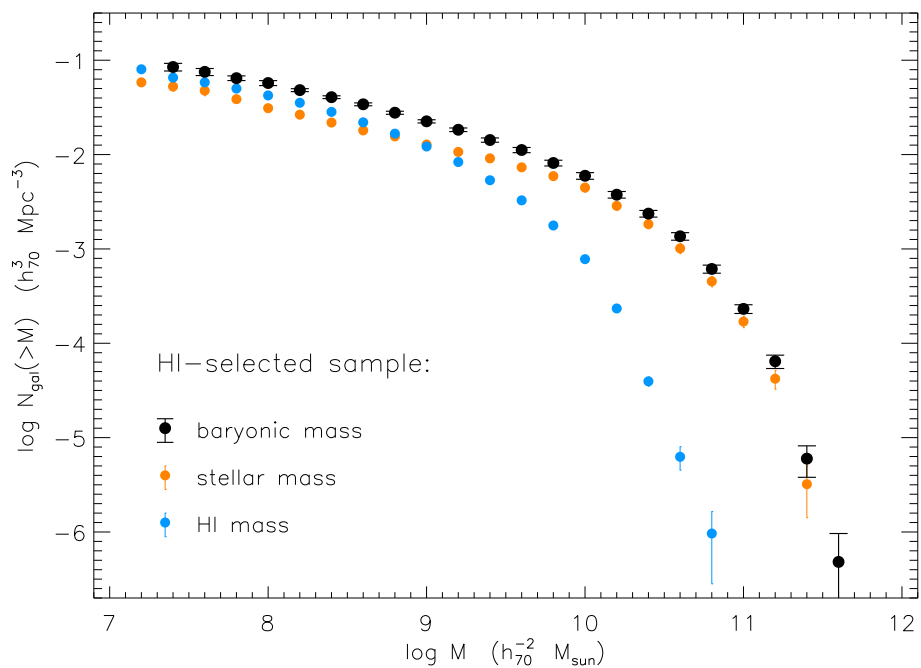}
\caption{ The \textit{cumulative} distributions of stellar mass (SMF, {\it gold symbols}), atomic hydrogen mass (HIMF, {\it cyan symbols})  and baryonic mass calculated as $M_b = M_\ast + 1.4 \: M_{HI}$ (BMF, {\it black symbols}), derived from the HI-selected galaxy sample. Error bars represent just the Poisson counting error assuming independent errors among different mass bins.  }
\label{fig:mfs_hi}
\end{figure}

\begin{figure}[htbp]
\includegraphics[scale=0.75]{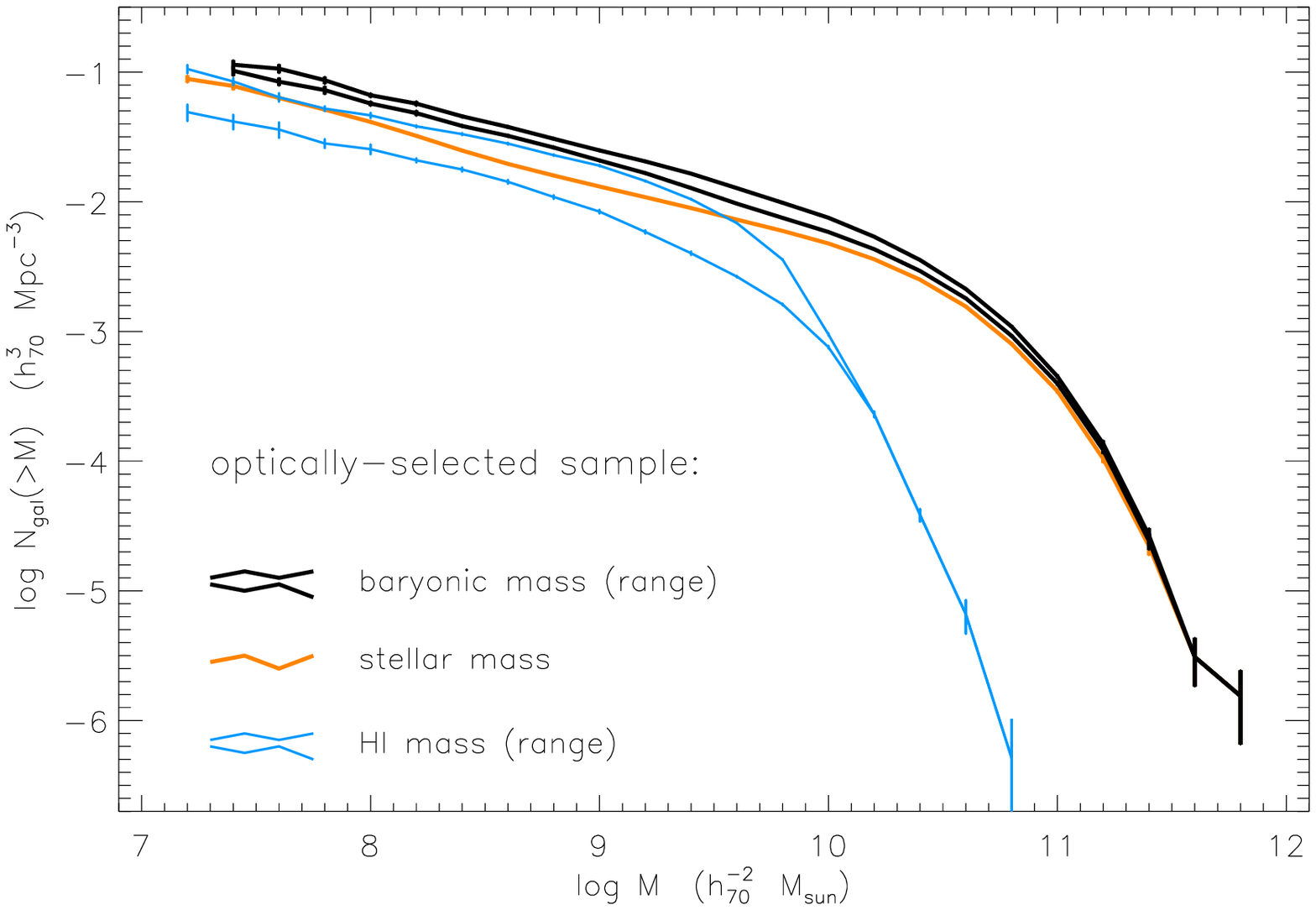}
\caption{The \textit{cumulative} distribution of stellar mass ({\it gold line}), atomic hydrogen mass ({\it cyan lines}) and baryonic mass calculated as $M_b = M_\ast + 1.4 \: M_{HI}$ ({\it black lines}), derived from the optically-selected galaxy sample. The atomic hydrogen and baryonic distributions are represented as allowed ranges, based on estimates of the minimum and maximum HI mass for galaxies that are not detected by ALFALFA (\{$M_{HI}^{min}, M_{HI}^{max}$\}, see \S \ref{sec:derived} for details). Error bars again represent just the Poisson counting error assuming independent errors among different mass bins. }
\label{fig:mfs_opt}
\end{figure}

Figure \ref{fig:mfs_hi} shows the cumulative distribution\footnotemark{} of stellar mass (SMF; {\it gold symbols}), HI mass (HIMF; {\it cyan symbols}) and baryonic mass (BMF; {\it black symbols}), derived from the HI-selected galaxy sample. The HI-selected BMF follows closely the HI-selected SMF at high masses, while at the low-mass end the contribution of the HIMF becomes dominant; this is because HI-selected galaxies become more gas-rich as their stellar mass decreases. Figure \ref{fig:mfs_opt} shows the same distributions (SMF, {\it gold line}; HIMF, {\it cyan lines}; BMF, {\it black lines}) derived from the optically-selected sample. Recall that, in the case of the optically-selected sample, a lower and upper limit of the HIMF and BMF are shown, since SDSS galaxies that are undetected by ALFALFA are assigned an upper and lower limit on their HI content, and therefore also on their baryonic mass (see \S \ref{sec:derived}). 

\footnotetext{ We show cumulative mass functions in Figs. \ref{fig:mfs_hi} \& \ref{fig:mfs_opt}, because the cumulative -and not the differential- distributions are directly related to the stellar and baryon galactic fractions computed in Sec. \ref{sec:bar_frac}. All other figures however show differential mass functions, which best display the details and errors of the distributions. }  


\begin{figure}[htbp]
\includegraphics[scale=0.65]{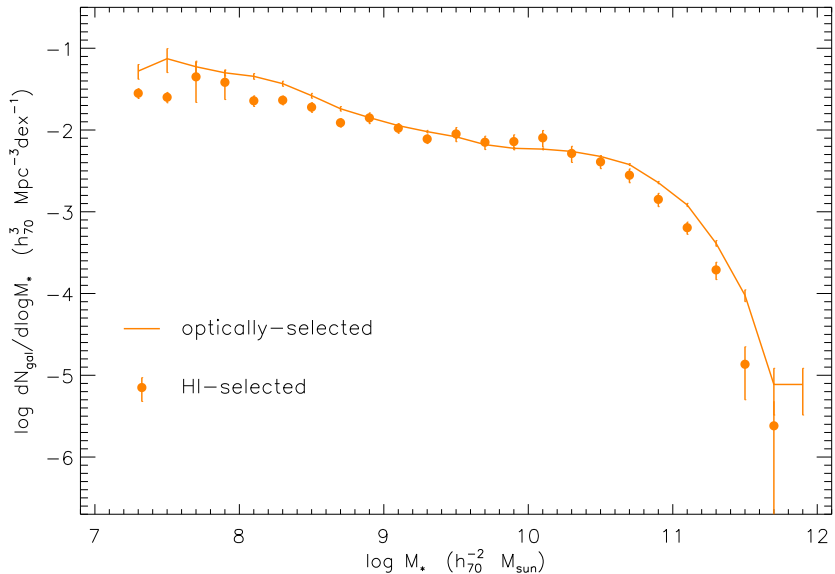}
\caption{ Comparison of the differential galactic stellar mass function (SMF) derived from the HI-selected ({\it gold symbols}) and optically-selected ({\it gold line}) galaxy samples. Error bars represent just the Poisson counting error on individual mass bins. The optically-selected SMF is systematically higher than the HI-selected SMF at the high-mass ($M_\ast \gtrsim 10^{11} \, M_\odot$) and low-mass ($M_\ast \lesssim 10^{8.5} \, M_\odot$) ends. This difference is mostly due to the bias of the HI-selected sample against the red galaxy population (see \S \ref{sec:results} for a detailed discussion).}
\label{fig:opt_vs_hi_gsmf}
\end{figure}

\begin{figure}[htbp]
\includegraphics[scale=0.65]{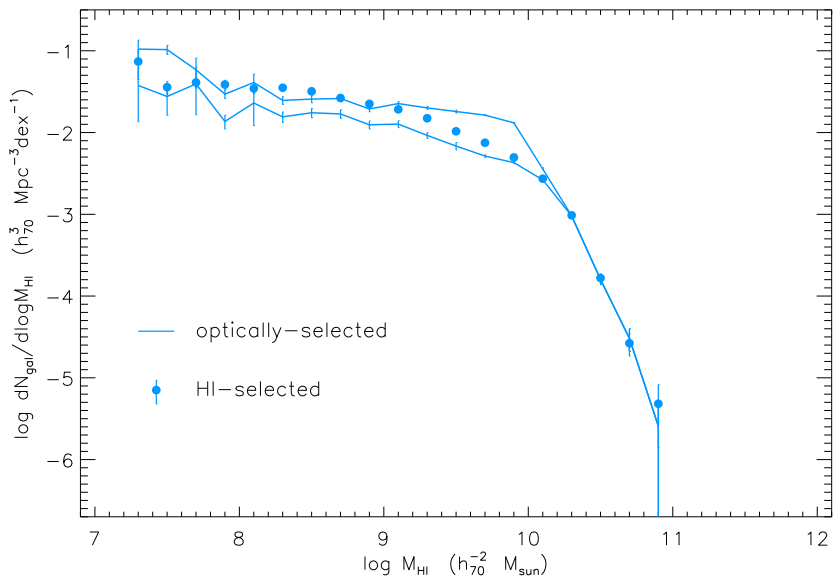}
\caption{ Comparison of the differential HI mass function (HIMF) derived from the HI-selected ({\it cyan symbols}) and optically-selected ({\it cyan lines}) galaxy samples. Error bars represent just the Poisson counting error on individual mass bins. The HIMFs derived from the two samples are mostly consistent, with the HI-selected HIMF having a slightly steeper low-mass end slope than that suggested by the optically-selected HIMF range. See \S \ref{sec:results} for a detailed discussion. }
\label{fig:opt_vs_hi_himf}
\end{figure}

\begin{figure}[htbp]
\includegraphics[scale=0.65]{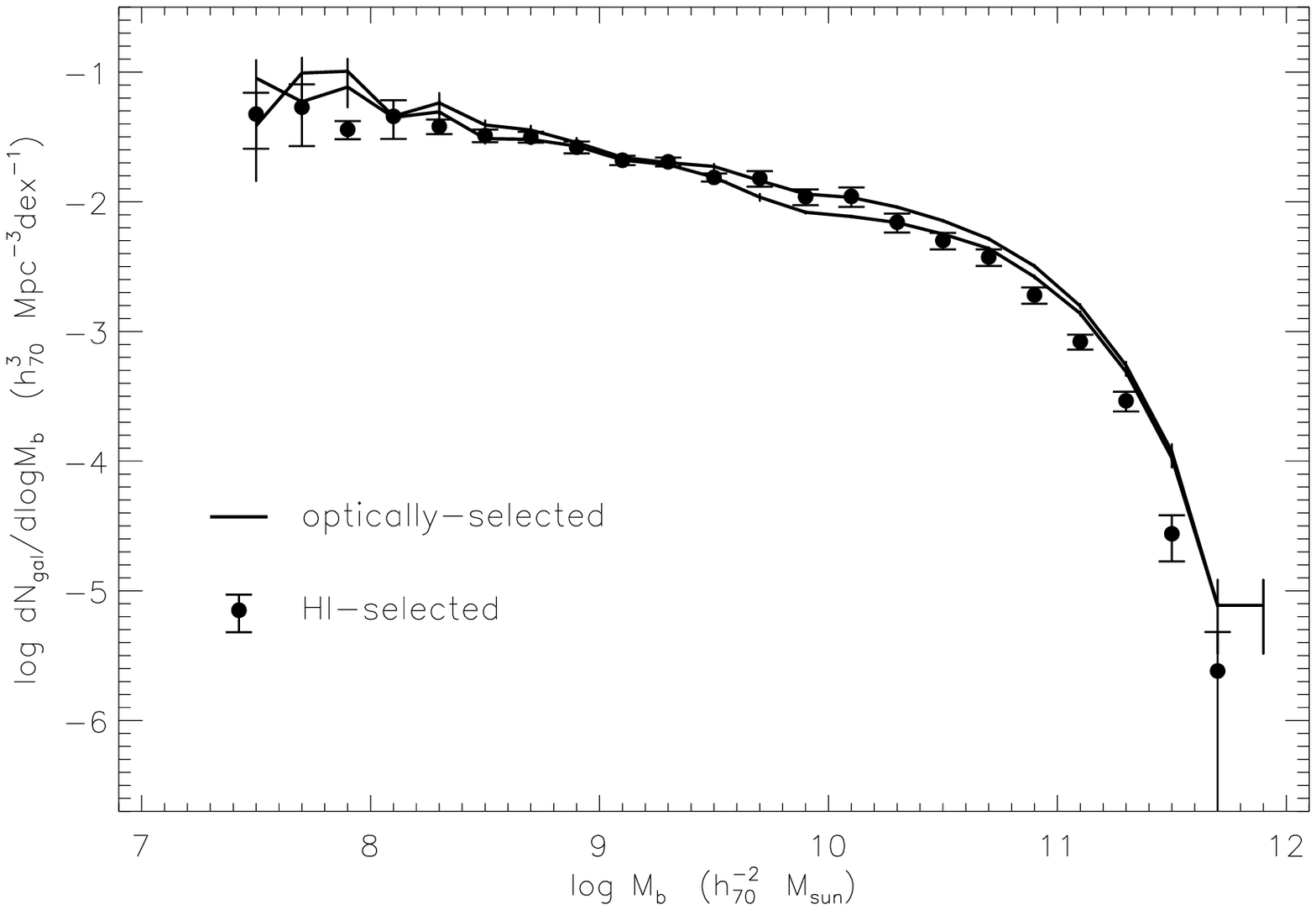}
\caption{ Comparison of the differential baryonic mass function (BMF) derived from the HI-selected ({\it black symbols}) and optically-selected ({\it black lines}) galaxy samples. Error bars represent just the Poisson counting error on individual mass bins. The optically-selected BMF is mostly consistent with the HI-selected BMF, except at the high-mass end. This is a direct result of the discrepancy between the optically-selected and HI-selected SMFs at high masses (Fig. \ref{fig:opt_vs_hi_gsmf}). See \S \ref{sec:results} for a detailed discussion. }
\label{fig:opt_vs_hi_bmf}
\end{figure}

In Figure \ref{fig:opt_vs_hi_gsmf} we compare the SMFs derived from the optically-selected and HI-selected samples. The two SMFs are consistent at intermediate masses, but the optically-selected SMF (gold line) is systematically higher at the high-mass and low-mass ends. At high masses the discrepancy is due to the bias of the HI-selected sample against the most massive galaxies, which are usually red passive systems. The discrepancy at the low-mass end is mostly due to the population of red-sequence dwarf galaxies in the nearby Virgo cluster\footnotemark{} that are undetected by ALFALFA; these are mostly dwarfs with early-type morphologies and very low HI content (see \citealp{Hallenbeck2012}, for example). On the other hand, the HI-selected and optically-selected HIMFs (Fig. \ref{fig:opt_vs_hi_himf}) are in good agreement with one another, with the HI-selected HIMF having a slightly steeper low-mass end slope than what suggested by the range of the optically-selected distribution. The two BMFs are mostly in agreement with one another, except at the high-mass end (a factor of $\approx$4 at $M_b = 10^{11.5} \, M_\odot$). This is a direct consequence of the discrepancy between the optically-selected and HI-selected stellar mass functions at high masses. Note that there is little difference between the two BMFs at low masses, which suggests that the low-mass end of the BMF has been measured robustly. 

\footnotetext{The presence of a massive cluster (Virgo cluster) at a distance of just 16.5 Mpc from the observer makes the ALFALFA survey volume different from an average cosmological volume. The effect of the presence of the Virgo cluster on the HIMF has been investigated by \citet[\S6.1]{Martin2010}, who found however only minor effects. More generally, the issue of cosmic variance regarding ALFALFA statistical distributions is discussed and quantified in \citet[\S4.3]{Papastergis2011}.}

\subsection{Comparison with other work   \label{sec:comparison}}

\begin{figure}[htbp]
\includegraphics[scale=0.75]{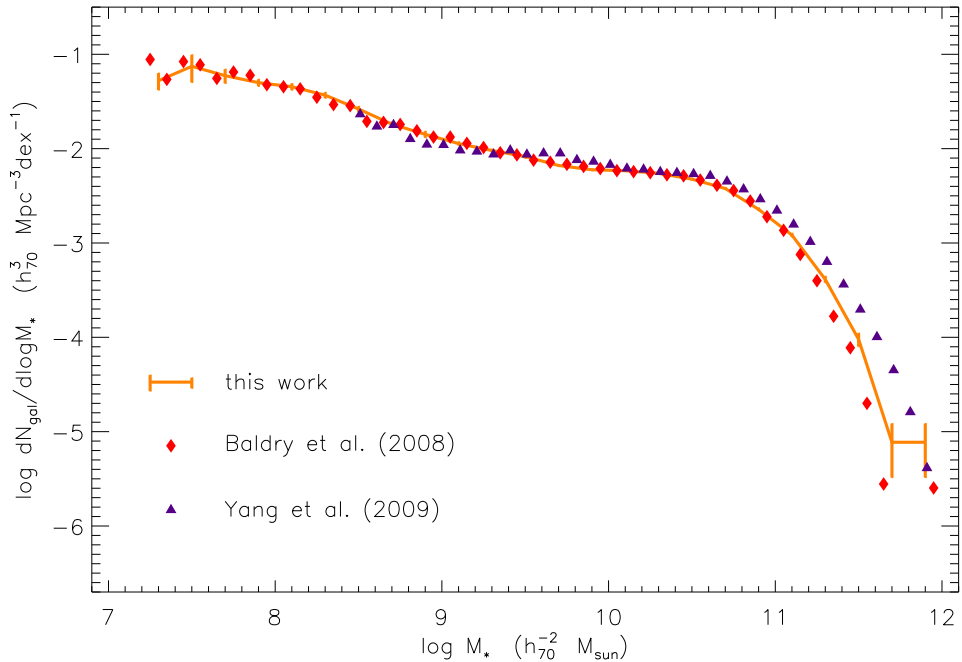}
\caption{The gold solid line with errorbars represents the differential SMF derived in this work from the optically-selected sample (same as gold line in Fig. \ref{fig:opt_vs_hi_gsmf}). The red diamonds correspond to the \citet{Baldry2008} SMF in the local universe ($z<0.06$), while the purple triangles correspond to the SMF of \citet{Yang2009}, extracted over a larger volume ($z<0.2$). Both the Baldry at el. and Yang et al. SMFs are based on the NYU-VAGC galaxy catalog. See \S \ref{sec:comparison} for a discussion of the comparison.}
\label{fig:gsmf_vs_other}
\end{figure}

Figure \ref{fig:gsmf_vs_other} compares the optically-selected SMF presented in this work (same as gold line in Fig. \ref{fig:opt_vs_hi_gsmf}) with the local-universe SMF of \citet{Baldry2008} and the \citet{Yang2009} SMF, which are both based on the New York University Value-Added Galaxy Catalog \citep[NYU-VAGC;][]{Blanton2005}\footnotemark{}. There is excellent agreement between the Baldry et al. SMF and our optically-selected SMF, especially at intermediate and low stellar masses ($M_\ast \lesssim 10^{11} M_\odot$). The deviations at high masses are due to the fact that stellar masses in Baldry et al. are calculated differently than in this work (see Sec. 3 of \citealp{Baldry2008} for details); note that a systematic difference of just 0.1 dex (26\%) in stellar mass would be enough to explain the observed difference in abundance at the high-mass end.     

\footnotetext{\texttt{http://sdss.physics.nyu.edu/vagc/}}

The \citet{Yang2009} SMF is systematically higher than our optically-selected SMF at high masses, and displays a more pronounced ``plateau'' at intermediate masses. It is not clear what the cause of the difference at the high-mass end of the distributions is, but it may relate to the fact that the Yang et al. SMF is extracted from a significantly larger volume than our measurement (the maximum redshift is $z = 0.2$ for the Yang et al. sample and $z = 0.05$ for the sample used in this work).  Moreover, Yang et al. use the prescription of \citet{Bell2003} to estimate stellar masses, which is based on the galactic $g-r$ color. As discussed in more detail in \S\ref{sec:stellar_est}, the use of different stellar mass estimators can significantly affect the shape of the measured SMF.

\begin{figure}[htbp]
\includegraphics[scale=0.75]{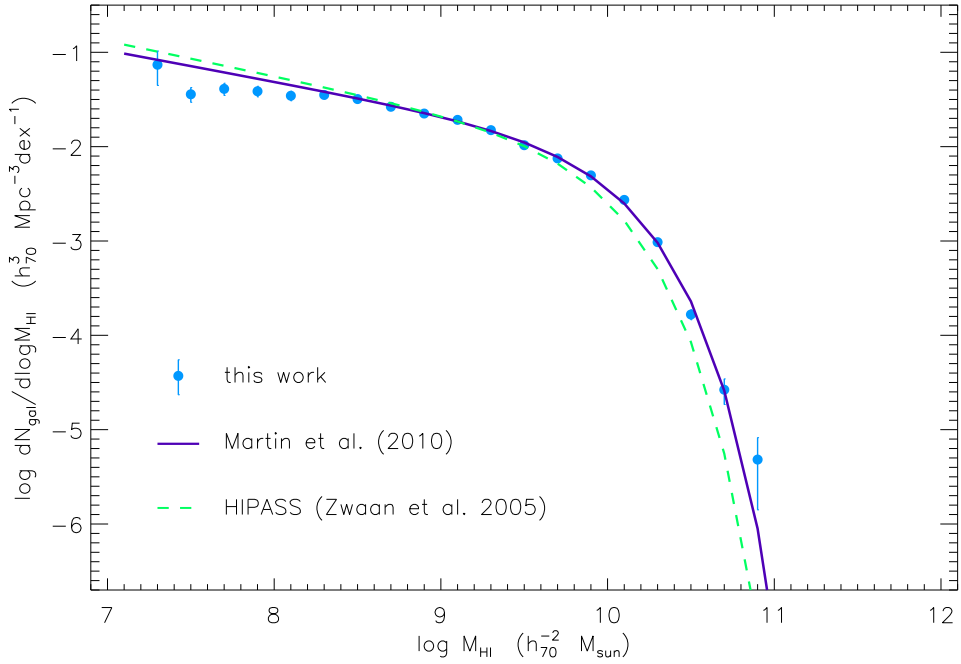}
\caption{The cyan symbols with error bars represent the differential HIMF derived in this work from the HI-selected sample (same as cyan symbols in Fig. \ref{fig:opt_vs_hi_himf}). The solid purple line corresponds to the Schechter function fit to the \citet{Martin2010} HIMF, which is based on the full \aforty \ catalog of ALFALFA sources and without any optical selection cuts. The green dashed line corresponds to the Schechter fit to the \citet{Zwaan2005} HIMF, based on 4315 galaxies detected by the HIPASS survey. See \S \ref{sec:comparison} for a discussion of the comparison. }
\label{fig:himf_vs_other}
\end{figure}

Figure \ref{fig:himf_vs_other} compares the HI-selected HIMF presented in this work (same as the cyan symbols in Fig. \ref{fig:opt_vs_hi_himf}) with the HIMF of \citet{Martin2010} derived from 10119 galaxies detected by ALFALFA over $\approx$2600 deg$^2$ of sky (purple solid line). There is excellent agreement at intermediate and large HI masses ($M_{HI} \gtrsim 10^{8.5} \, M_\odot$) between the \citet{Martin2010} HIMF and the HIMF derived in this work, while at lower masses the \citet{Martin2010} HIMF is slightly steeper than ours. This disagreement may be due to the set of additional optical requirements imposed on our HI-selected sample. As argued in \S\ref{sec:hi_selected}, these requirements are expected to reduce the number of low-mass systems in the sample and therefore decrease the inferred space density at the low-mass end. The dashed green line represents the HIMF of \citet{Zwaan2005} based on 4315 sources detected by the HI Parkes All-Sky Survey (HIPASS) over the whole southern celestial hemisphere ($\approx$ 29000 deg$^2$). There is disagreement between both ALFALFA-based HIMFs and the HIPASS-based HIMF of \citet{Zwaan2005}, at the high-mass end. As argued in \citet{Martin2010}, the higher sensitivity of the ALFALFA survey compared to HIPASS, which enables ALFALFA to detect HI-massive systems over a larger volume, should give a statistical advantage to the ALFALFA survey in determining the high-mass end of the HIMF. However, the difference is too large to be explained by counting statistics or cosmic variance (e.g. according to the estimates of \citealp{Somerville2004} or \citealp{DR2010}). On the other hand, due to the exponential drop-off of the HIMF at high masses, a flux calibration difference of as low as 0.1 dex could give rise to a similar discrepancy.

\section{Uncertainties \& systematics        \label{sec:biases}} 

\subsection{Stellar mass estimator   \label{sec:stellar_est}}

\begin{figure}[htbp]
\includegraphics[scale=0.75]{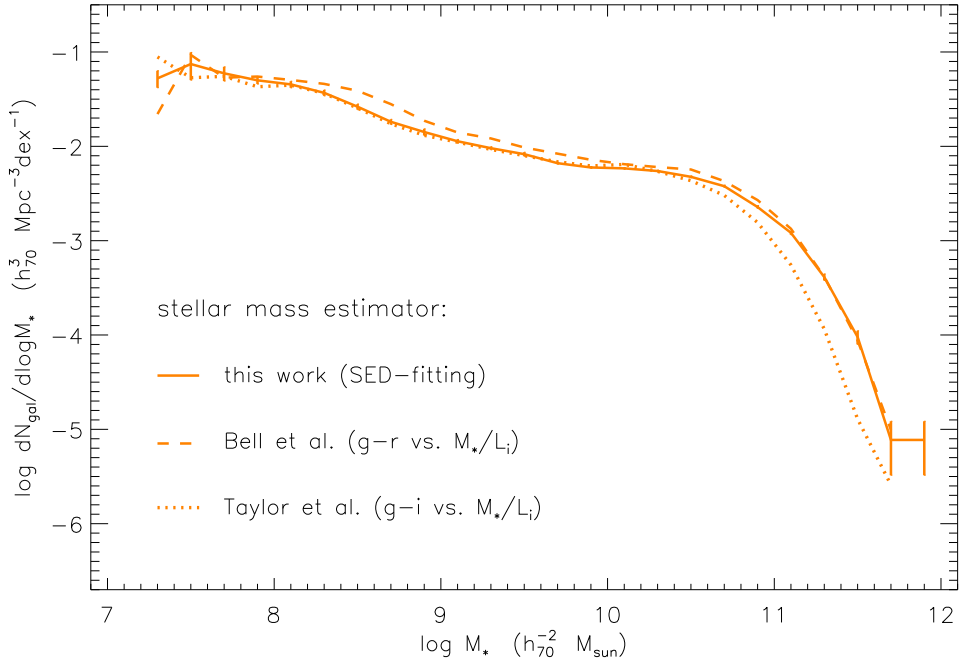}
\caption{The gold solid line with error bars represents the differential SMF derived from the optically-selected sample in this work, using stellar masses based on SED-fitting \citep{Huang2012a}. The gold dashed line represents the SMF computed from the same sample but using stellar masses derived from the galactic $g-r$ color and the $i$-band luminosity, according to the widely used \citet{Bell2003} calibration. The gold dotted line represents the SMF computed using stellar masses derived from the galactic $g-i$ color and the $i$-band luminosity, according to the the more recent calibration of \citet{Taylor2011}. See \S\ref{sec:stellar_est} for a discussion of the comparison. }
\label{fig:sed_vs_color}
\end{figure}

A variety of methods exist to estimate stellar masses from spectra or broadband photometric measurements of galaxies \citep[][to name a few]{Kauffmann2003, Bell2003, Brinchmann2004, Glazebrook2004, Gallazzi2005, Panter2007, Salim2007}. Most methods rely on comparing the actual galactic emission to the light output of a set of galactic stellar population models. The models that most closely reproduce the observed data are then used to estimate the galactic properties of interest (e.g. stellar mass, present star formation rate, internal extinction etc.); it is therefore very important to consider model stellar populations which span as large a range of physical parameters as the galaxies in the sample being studied. Inferred galactic properties depend not only on the particular type of data employed by each method (e.g. spectroscopy vs. broadband photometry or optical vs. near-infrared photometry), but also by differences in the way in which the model stellar populations are constructed. This means that different methods can yield different estimates of a galactic property even when the \textit{same} observational measurements are used. For example, \citet{Maraston2012} find that unbiased stellar masses can only be recovered when the true star formation history (SFH) of a galaxy is known. In practice however a restricted set of SFHs is considered (often in the form of a parametrized function), which may introduce systematics for galaxies with SFHs that are not well described by the assumed general form. Additional complications can be introduced by the different treatment of dust reddening among different models. In general, stellar mass estimates can differ systematically by as much as 0.3 dex, while for individual galaxies the scatter can be as large as 0.6 dex \citep{Maraston2012}.

\begin{figure}[htbp]
\includegraphics[scale=0.7]{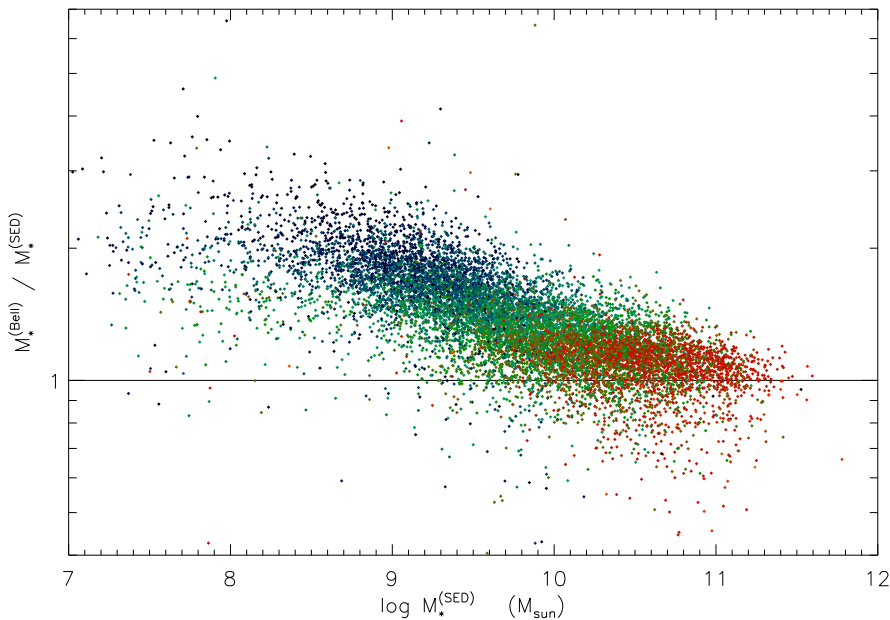}
\caption{ A galaxy-by-galaxy comparison of stellar masses derived from SED fitting of the SDSS $u,g,r,i,z$ bands used in this work (see \S \ref{sec:derived}), and those derived from the galactic $g-r$ color and the $i$-band luminosity according to the widely used \citet{Bell2003} calibration. Each datapoint corresponds to a galaxy in the optically-selected sample, and the symbol color represents the g-r color of the galaxy. The two stellar mass estimates agree fairly well for red passive galaxies, while for blue star-forming galaxies Bell et al. masses are systematically larger by up to a factor of $\approx$2 (see \S\ref{sec:stellar_est} for a detailed discussion). }
\label{fig:bell_vs_sed}
\end{figure}

Here we compare the optically-selected SMF presented in this work (computed from stellar masses derived from SED-fitting, see \S\ref{sec:derived}) with the SMF obtained using stellar masses derived from a single galactic color, according to the widely-used \citet{Bell2003} calibration as well as the more recent calibration of \citet{Taylor2011}. More specifically, we compute Bell et al. masses by multiplying the $i$-band luminosity of each galaxy by a mass-to-light ratio inferred from its $g-r$ color\footnotemark{}. We choose this particular combination of bands because it is relatively immune to contamination of galactic spectra by bright nebular emission lines \citep{West2009}. We use a similar procedure to calculate Taylor et al. masses, by using their calibration of $i$-band mass-to-light ratio versus $g-i$ color. \citet{Taylor2011} argue that using $g-i$ colors best constrains the galactic stellar mass estimates. 

Figure \ref{fig:sed_vs_color} shows the impact that different methods of estimating stellar mass have on the measurement of the SMF. When Bell et al. stellar masses are used (\textit{gold dashed line}), the SMF becomes systematically higher at low and intermediate masses ($M_\ast \lesssim 10^{11} \, M_\odot$), while it remains mostly unchanged at the high-mass end. The reason for this pattern becomes evident in Fig. \ref{fig:bell_vs_sed}, where we see that Bell et al. masses agree fairly well with the masses derived in this work for red passive galaxies, but are systematically larger (by up to a factor of $\approx$2) for blue star-forming galaxies. \citet{Huang2012a} argue that the difference can be primarily attributed to the fact that the stellar population models used for the Bell et al. calibration do not consider ``bursty'' star formation histories which are typical of low-mass galaxies with blue colors. This leads to systematically older stellar populations for blue galaxies according to the Bell et al. method, which in turn results in systematically higher stellar mass estimates. Note that including models with bursty SFHs in a stellar population library does not by itself guarantee a correct estimate of stellar mass; overestimating the effect of bursts would result in systematically low stellar masses for blue galaxies. Conversely, when Taylor et al. masses are used (gold dotted line), the SMF becomes systematically lower at the high-mass end, while it is mostly unchanged at low and intermediate masses. Again, this is a result of the fact that Taylor et al. masses agree well with the SED-fitting masses used in this work for blue star-forming galaxies, but are systematically lower (by up to a factor of $\approx$1.4) for red passive galaxies.

\footnotetext{We use SDSS colors, computed from Galactic extinction-corrected model magnitudes (\texttt{modelmag}), to calculate mass-to-light ratios in the $i$-band. $i$-band luminosities are then calculated from the $i$-band Petrosian magnitudes reported in SDSS (\texttt{petromag}), corrected for Galactic extinction according to the values listed in the SDSS database. The solar absolute magnitude in the $i$-band is taken to be $M_{\odot,i} = 4.57$.}


\subsection{Distance uncertainties  \label{sec:dist_unc}}

\begin{figure}[htbp]
\includegraphics[scale=0.75]{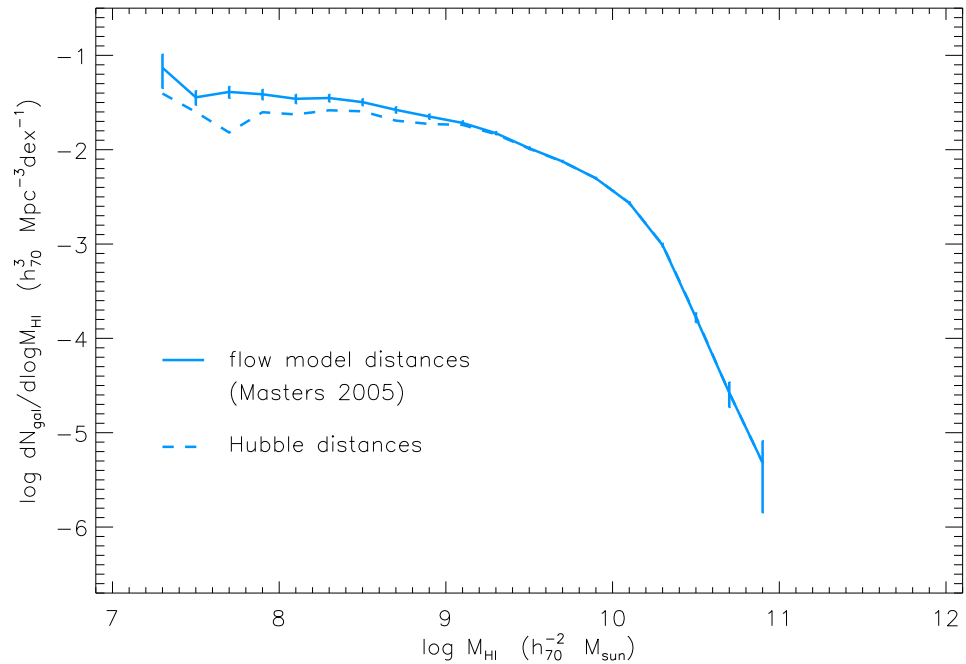}
\caption{The cyan solid line with error bars represents the differential HIMF derived from the HI-selected sample in this work, using flow model distances \citep{Masters2005} for most nearby galaxies. The cyan dashed line represents the HIMF computed from the same sample but using simple Hubble distances for all galaxies. Distance uncertainties affect primarily the mass estimates of nearby galaxies and so the main effect is a change of the low-mass end slope of the distribution (see \S\ref{sec:dist_unc} for further discussion).  }
\label{fig:hubble_vs_flow}
\end{figure}

Stellar, HI and baryonic masses are distance-dependent quantities and hence their statistical distributions are affected by distance uncertainties. This is particularly true for the low-mass end of the distributions, which is determined by the properties of nearby galaxies; neglecting the peculiar velocity of some of these objects can cause fractional distance and mass errors of order $\approx 100\%$, especially in a volume with complex large-scale structure such as the one surveyed by ALFALFA (see Fig. \ref{fig:coneplot}). 

For this reason, nearby galaxies ($v_{CMB} \leqslant 6000$ \kms) in the \aforty \ catalog are assigned distances based on a parametric peculiar velocity flow model \citep{Masters2005}, and only more distant galaxies ($v_{CMB} > 6000$ \kms) are assigned simple Hubble distances according to their CMB recessional velocity. The \citet{Masters2005} flow model includes a dipole and a quadrupole component (local group bulk motion \& asymmetric expansion) and two local attractors (Virgo cluster \& Great Attractor), and is calibrated against the SFI++ catalog of galaxies with redshift and independent distance distance information (from Tully-Fisher). The residuals are then attributed to random thermal motions, which are estimated to have a magnitude of $\sigma_{local} = 160$ \kms. In addition, distances reported in the \aforty \ catalog take into account known group and cluster membership as well as primary distance information published in the literature. This latter information is not available for the majority of the galaxies in our optically-selected sample (which are not included in \aforty), and we only make an attempt to assign all probable Virgo members to the Virgo cluster distance ($D = 16.5 \; \mathrm{Mpc}$).

Here we re-evaluate the HIMF for our HI-selected sample using uniformly Hubble distances for all galaxies, in order to illustrate the impact of the distance assignment scheme on the derived distributions. Figure  \ref{fig:hubble_vs_flow} shows that the HIMF computed using Hubble distances (cyan dashed line) has a much shallower low-mass end slope compared to the HIMF presented in this work, which uses flow model distances for most nearby galaxies (cyan solid line). This result is in agreement with the work of \citet{Masters2004}, who find that neglecting the local peculiar velocity field when estimating distances to nearby galaxies in the ALFALFA volume will lead to a systematically shallower low-mass end slope for the HIMF.

\subsection{Molecular \& ionized gas   \label{sec:h2}}

\begin{figure}[htbp]
\includegraphics[scale=0.75]{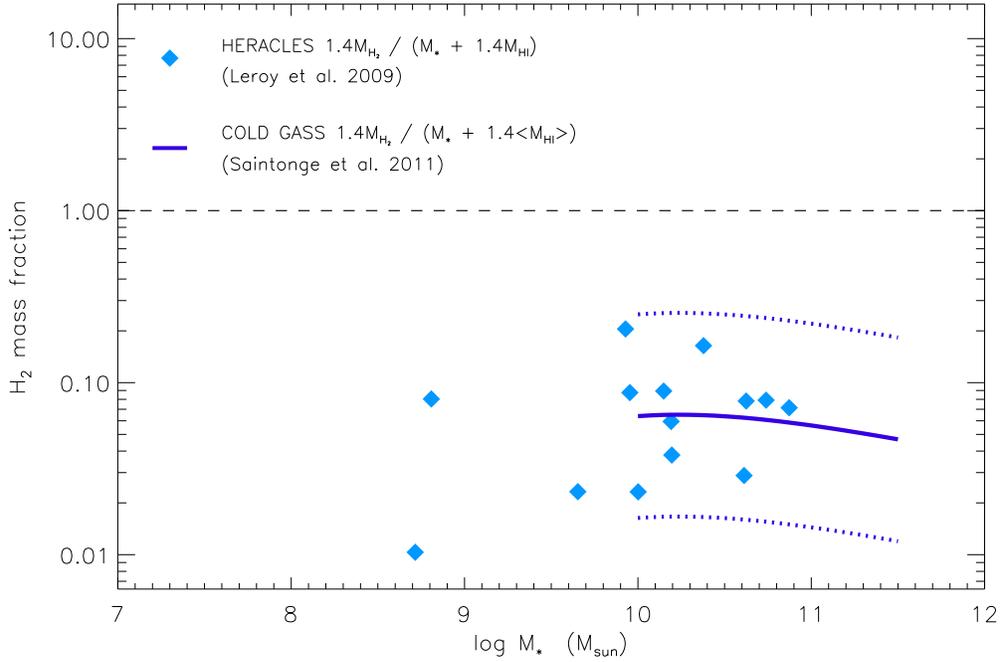}
\caption{Ratio of molecular gas mass (corrected for the abundance of helium) to the ``baryonic mass'' computed as $M_b = M_\ast + 1.4 \, M_{HI}$. Cyan diamonds represent 14 galaxies of the HERACLES survey \citep{Leroy2009} detected in CO line emission. The solid and dotted blue lines represent the average and $2\sigma$ scatter of the distribution found for 125 CO detected galaxies in the COLD GASS survey \citep{Saintonge2011}. Both samples show that, at least for $M_\ast > 10^{8.5} M_\odot$, molecular gas is almost always a subdominant mass component (see \S\ref{sec:h2} for details).}
\label{fig:h2_mass_frac}
\end{figure}

Throughout this article we use the term ``baryonic mass'' to refer to the sum of the stellar and atomic gas mass ($M_b = M_\ast + 1.4 \, M_{HI}$), a convention that is common in the literature. This definition however excludes a number of baryonic components that are definitely present in galaxies, most notably molecular and  ionized -warm or hot- gas. 


Figure \ref{fig:h2_mass_frac} displays the fraction of molecular hydrogen (\htwo) mass (accounting for helium) to ``baryonic mass'' as defined above, as a function of stellar mass. The cyan diamonds represent 14 galaxies from the HERACLES survey \citep{Leroy2009} with \htwo \ masses measured from interferometric CO line observations (using a fixed $\alpha_{CO}$ conversion factor). Over the probed stellar mass range ($M_\ast \gtrsim 10^{8.5} M_\odot$), molecular gas rarely contributes more than 10\% of the ``baryonic mass''. The same conclusion is reached when \htwo \ mass measurements from the COLD GASS survey are considered \citep{Saintonge2011}. The blue solid and dotted lines show the mean and $2\sigma$ scatter of the relation between the molecular and ``baryonic'' mass components, based on 125 galaxies detected in CO emission with the IRAM 30m telescope. In this latter case we have estimated the atomic hydrogen mass of galaxies indirectly, using the average $M_{HI}/M_\ast$ vs. $M_\ast$ relation of the COLD GASS parent sample \citep{Catinella2010}. Again, over the stellar mass range probed by the survey ($M_\ast = 10^{10} - 10^{11.5} \, M_\odot$) molecular gas is always a subdominant mass component. At lower stellar masses there is large uncertainty on the fractional contribution of \htwo, as it is not precisely known how well the galactic CO emission traces molecular hydrogen mass. In particular, the $\alpha_{CO}$ conversion factor may vary by about an order of magnitude as we consider less luminous and more metal poor late-type galaxies \citep[e.g.][]{Boselli2002}.   

Determining the contribution of ionized gas to the total baryonic mass budget of galaxies is much more challenging. For example \citet{Reynolds2004} argue that warm ionized hydrogen (HII) may amount to about 1/3 of the mass of atomic hydrogen (HI) in the disk of the Milky Way. If the ratio of ionized-to-neutral hydrogen mass ($f_\frac{HII}{HI}$) were fairly constant among galaxies, then the baryonic mass of a galaxy would be given by the expression $M_b = M_\ast + 1.4 \, (1+f_\frac{HII}{HI})M_{HI}$. If $f_\frac{HII}{HI} \approx 0.3$, then the peak value of the $\eta_b$ - $M_h$ relation (see Fig. \ref{fig:am2}) would shift to lower halo mass and the peak value would slightly increase. However, since the precise value and scatter of $f_\frac{HII}{HI}$ is not well constrained -and its dependence on galaxy size is not known- we choose not to include the contribution of warm ionized gas in the calculation of $M_b$.

Assessing the contribution of the hot ionized medium (HIM) to the total baryonic mass of a galaxy is even more challenging. The coronal HIM may be the dominant baryonic mass component in galactic halos, especially in massive ellipticals. Determining the mass contribution of the HIM for less massive galaxies however is observationally challenging. In any case, the tightness of the ``baryonic Tully-Fisher relation'' when computed just from the stellar and HI mass \citep[e.g.][]{McGaugh2012, Hall2012} implies that the HIM never dominates the total baryonic mass budget of late-type galaxies, at least within the extent of the galactic HI disk.


\section{The stellar \& gas content of DM haloes     \label{sec:bar_frac}}

\subsection{The abundance matching method and its application   \label{sec:am}}

Let $N_{\rm gal}(M_b)$ be the cosmic number density of galaxies with baryon mass greater than $M_b$ and let $N_h(M_h)$ be the cosmic number density of haloes with mass greater than $M_h$. The fundamental assumption of the abundance matching method
(\citealp{MarinoniHudson2002}; \citealp{ValeOstriker2003}; also see \citealp{Behroozi2010} for a review) is that $M_b$ is a monotonically increasing function of $M_h$. With this assumption, $M_b \, (M_h)$ can be determined by solving the equation 

\begin{equation}
\label{eq:am}
N_{\rm gal}(M_b)=N_h(M_h).
\end{equation}

\noindent
In reality, the baryon content of a halo will depend not only on its mass but also on other parameters, such as its formation history. As a result, a scatter in the distribution of $M_b$ at a given $M_h$ is expected. Neglecting the scatter is, nevertheless, justifiable because the aim of abundance matching is precisely to determine the average value of $M_b$ within a halo of mass $M_h$. 


We evaluate the right hand side of equation~\ref{eq:am}, using a halo mass function extracted  from one of the cosmological N-body simulations of the Horizon Project\footnotemark{}. The simulation was run with a public version of the GADGET code \citep{Springel2005}, and uses $1024^3$ particles of mass $m_p \sim 8.5 \ 10^7 \ M_\odot$ to simulate the formation and evolution of DM structures in a comoving volume of 100 $h^{-1}$ Mpc on a side. It assumes a cosmology and initial conditions which are consistent with Wilkinson Microwave Anisotropy Probe (WMAP) third year results \citep{Spergel2007}, namely $h = 0.73$, $\Omega_\Lambda=0.76$, $\Omega_m =0.24$, and $\sigma_8 =0.76$. The identification of  DM haloes and sub-haloes was done with the adaptaHOP algorithm presented in \citet{Aubert2004}. Haloes are identified as groups of particles above a threshold over-density of 80 times the mean density of the universe, which corresponds to a mean overdensity contrast of about 200. The identification of subhaloes within haloes is done using the  method described in \citet{Tweed2009}. We only keep haloes and sub-haloes with more than 20 particles, i.e. we introduce a minimum halo mass of $M_h \approx 1.7 \ 10^9 \ M_\odot$.

\footnotetext{\texttt{http://www.projet-horizon.fr}}

\begin{figure}[htbp]
\includegraphics[scale=0.7]{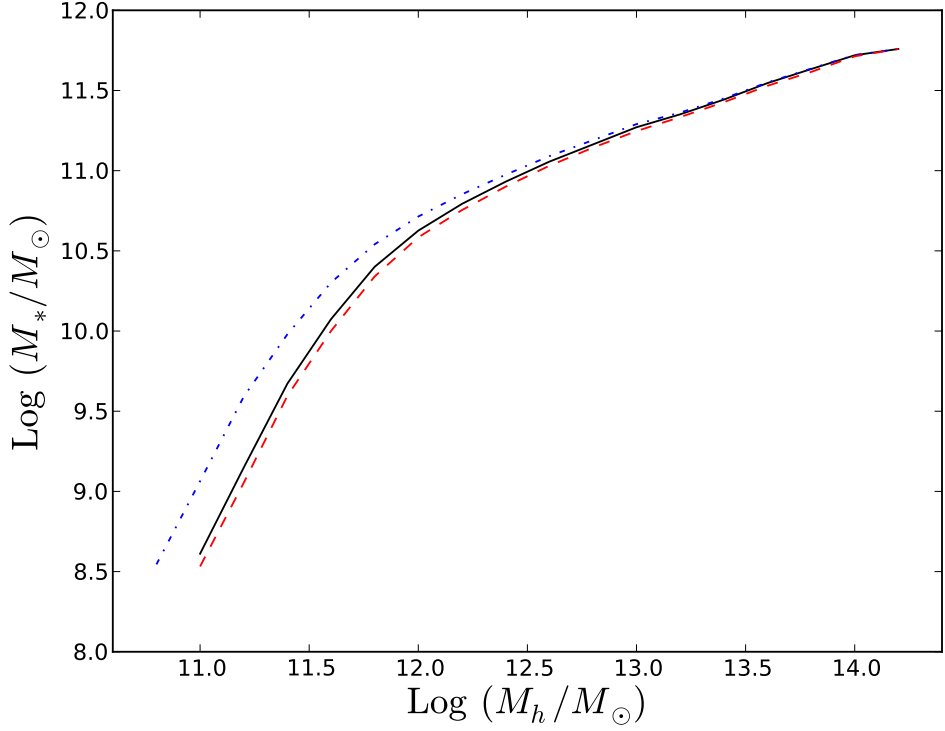}
\caption{Effects of the choice of stellar and halo mass functions on the $M_\ast$ - $M_h$ relation. The black solid line shows the $M_\ast$ - $M_h$ relation obtained by matching the stellar mass function of central galaxies to the mass function of haloes excluding subhaloes. This matching should reproduce the ``true'' relation for central galaxies. The blue dot-dashed line is the relation obtained  by matching the total galaxy stellar mass function, including central and satellite galaxies, to the mass function of haloes excluding subhaloes. The red dashed line is the relation obtained by matching the total galaxy stellar mass function with the mass function of haloes including subhaloes. Note that, unlike all other figures, this figure uses the stellar mass functions of \citet{Yang2009}, who have separately measured the SMF for central and satellite galaxies. }
\label{censat}
\end{figure}

\begin{figure}[htbp]
\includegraphics[scale=0.7]{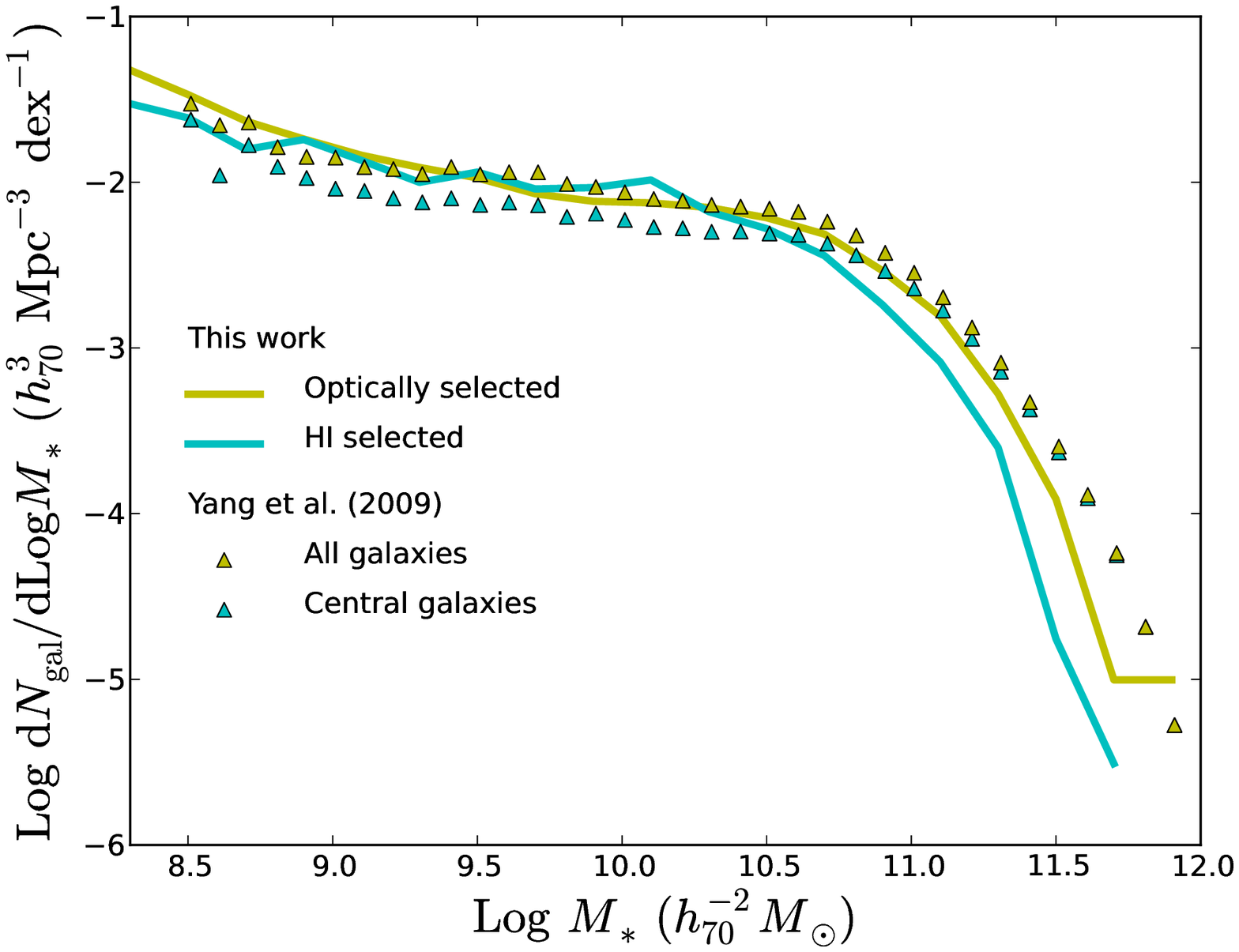}
\caption{The differential SMFs derived from our optically selected sample (yellow curve) and our HI-selected sample (cyan curve) compared to the results of \citet{Yang2009} for the total SMF (yellow symbols) and the SMF for central galaxies only (cyan symbols). }
\label{censatbis}
\end{figure}

An important issue for our analysis is determining whether subhaloes should be included or excluded in the calculation of the HMF used in the abundance matching procedure. Let $M_\ast^c(M_h)$ be the stellar mass of the central galaxy in a halo of mass $M_h$ and let $M_\ast^s(M_h)$ be the stellar mass of the satellite galaxy in a subhalo of mass $M_h$. If subhaloes are excluded, then we implicitly assume that all galaxies in our samples are central galaxies, that is, $M_\ast^s(M_h)=0$. On the other hand, if subhaloes are included, we implicitly assume that the $M_b$ - $M_h$ relation is the same for both central and satellite galaxies, that is, $M_\ast^s(M_h)=M_\ast^c(M_h)$. 

In order to understand how much this choice could affect our results, we consider the stellar mass functions of \citet{Yang2009}, for which separate distributions for central and satellite galaxies have been presented. This allows us to calculate the $M_\ast$ - $M_h$ relation in three ways: firstly, we can match the SMF of central galaxies with the halo mass function excluding subhaloes; this method will give us the correct $M_\ast$ - $M_h$ for central galaxies, shown by the black solid line in figure~\ref{censat}. Then, we consider the total SMF of central plus satellite galaxies and match it with the halo mass function excluding subhaloes. This is equivalent to assuming that all galaxies are central and overestimates the $M_\ast$ - $M_h$ for central galaxies (blue dot-dashed line in Fig.~\ref{censat}). Finally, we can match the total SMF of central plus satellite galaxies with the total halo mass function including subhaloes. The result, shown by the red dashed line in Figure~\ref{censat}, lies below the black solid line because it is effectively a weighted average of the relation for the dominant central galaxy population (the black solid line) and the relation for the satellite population, which has lower $M_\ast$ for a given $M_h$. Quantitative comparisons of  $M_\ast^s(M_h)$ and $M_\ast^c(M_h)$ have received the attention of much recent literature (see e.g. Cattaneo et al. 2012, submitted, \citealp{Rodriguez2012}, \citealp{Reddick2012}). 

As we do not distinguish between central and satellite galaxies in the samples used in this work, we shall choose the third method (the one that corresponds to the red dashed line) as our best estimator of the $M_\ast$ - $M_h$ relation and the $M_b$ - $M_h$ relation. This choice will introduce some systematic bias which is, nonetheless, smaller than the typical uncertainty involved in the determination of the $M_\ast$ - $M_h$ relation.

We also considered whether abundance matching our HI-selected SMF with the HMF excluding subhaloes would give consistent results with our fiducial abundance matching result, obtained by matching our optically-selected SMF and  the HMF including subhaloes. Physically, this consideration was motivated by the fact that satellite galaxies tend to be redder than central galaxies of the same mass \citep[e.g.][]{Weinmann2006}, and so HI-selection may be equivalent to the exclusion of satellite galaxies from an observational sample. However, the comparison in Fig.~\ref{censatbis} of our HI-selected SMF with the SMF for central galaxies of \citet{Yang2009} shows that this argument may not be valid. We note that the comparison between the Yang et al. SMF for central galaxies and our HI-selected SMF is subject to observational systematics, such as the  distance assignment scheme or the stellar mass estimator; for example using Hubble distances for the galaxies in our HI-selected sample (see Fig.~\ref{fig:hubble_vs_flow}) would bring the two distributions in fair agreement.      

%
%
%

\subsection{Results}

\begin{figure}[htbp]
\includegraphics[scale=0.7]{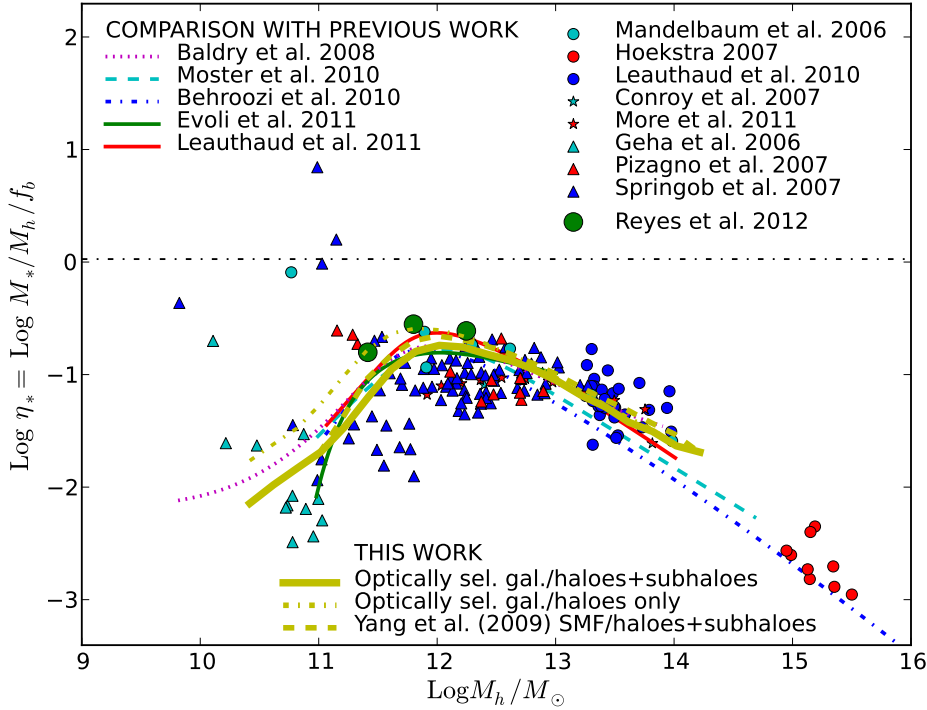}
\caption{The ratio of galactic stellar mass to halo mass as a function of host halo mass ($M_\ast / M_h - M_h$ relation). The thick yellow line shows our main result, obtained from abundance matching the stellar mass function of our optically-selected sample with the halo mass function including subhaloes. The yellow dashed and dash-dotted lines correspond to variations of our main result, obtained by considering the \citet{Yang2009} SMF and excluding subhaloes from the HMF respectively, and are shown to illustrate uncertainties. The magenta dotted, cyan dashed, blue dot-dashed, green solid and red solid lines correspond to the abundance matching results of \citet{Baldry2008}, \citet{Moster2010},  \citet{Behroozi2010}, \citet{Evoli2011} and \citet{Leauthaud2012}, respectively. The big green circles are the results of a stacked weak lensing study of SDSS galaxies by \citet{Reyes2012}. All other data points refer to measurements for individual galaxies: the small circles correspond to galaxies with halo mass measurements from weak lensing studies (\citealp{Mandelbaum2006}: cyan circles; \citealp{Hoekstra2007}: red circles; \citealp{Leauthaud2010}: blue circles). The star symbols show galaxies in which the halo mass was determined from studies of stellar dynamics (\citealp{Conroy2007}: cyan stars; \citealp{More2011}: red stars). The triangles show galaxies with halo masses determined from the disc rotation speed ( \citealp{Geha2006}: cyan triangles; \citealp{Pizagno2007}: red triangles; \citealp{Springob2007}: blue triangles). The dotted-dashed horizontal line shows the cosmic baryon fraction $f_b \approx 0.16$.  }
\label{fig:am1}
\end{figure}


Our abundance matching analysis produces two main results:

\begin{enumerate}
 
\item We determine  $M_\ast/M_h$ as a function of  $M_h$ by comparing the galaxy stellar mass function (SMF) from our optical sample to the total halo mass function (including subhaloes).
\label{item1}

\item We determine $M_b/M_h$ as a function of $M_h$ by comparing the baryonic mass function (BMF) from our optically selected sample to the total halo mass function (including subhaloes).
\label{item2}

\end{enumerate}

The latter relation is the focus of this paper, but we first discuss the former because the results can be compared to an extensive literature of previous studies. The consistency of our findings with previous work on point \ref{item1} boosts our confidence that our conclusions on point \ref{item2} are robust.

Figure~\ref{fig:am1} shows our result for $M_\ast/M_h$ as a function of  $M_h$, obtained from our optically selected sample (gold thick solid line). Our analysis extends to halo masses as low as $M_h \approx 10^{10.5} \, M_\odot$, since both our optically-selected and HI-selected samples probe stellar masses down to $M_\ast \approx 10^7 \,  M_\odot$. At the high mass end, our relation stops at $M_h \approx 10^{14}\,M_\odot$ because our galactic samples are drawn from a relatively small volume, and are not appropriate to measure the abundance of the most massive galaxies and clusters. We find good agreement with previous estimates from abundance matching obtained in the same manner (\citealp{Moster2010}: cyan dashed line; \citealp{Behroozi2010}: blue dotted-dashed line; \citealp{Evoli2011}: green solid line; \citealp{Leauthaud2012}: red solid line; \citealp{Baldry2008}\footnotemark{}: purple dotted line). The thick yellow dash-dotted and dashed lines are shown to illustrate the systematics introduced by the choice of HMF and SMF: the former represents the abundance matching result when our fiducial SMF is matched the HMF excluding subhaloes; the latter is the result of matching the Yang et al. SMF with our fiducial HMF, which includes subhaloes.

\footnotetext{The \citet{Baldry2008} abundance matching result uses a ``galactic'' halo mass function by \citet{Shankar2006}.}

We also compare our average $M_\ast$ - $M_h$ relation with values measured for individual galaxies. The small cyan, red and blue circles correspond to galaxies with measurements of their halo mass $M_h$ from weak lensing \citep{Mandelbaum2006, Hoekstra2007, Leauthaud2010}. The star symbols correspond to galaxies for which $M_h$ was estimated form stellar dynamics \citep{Conroy2007, More2011}. The triangles correspond to disc galaxies for which $M_h$ was determined from their rotation speed \citep{Geha2006, Pizagno2007, Springob2007}. We remark that, while results for individual galaxies have large scatter, they seem to be systematically lower than any of the relations inferred from abundance matching (at least over the halo mass range $M_h = 10^{11} - 10^{12}  \, M_\odot$). Furthermore, the halo mass for which $M_\ast / M_h$ has a maximum appears to be higher when inferred from measurements of individual galaxies compared to the value derived from abundance matching: in the first case the peak is at $M_h \approx 10^{12.5}\,M_\odot$, while in the second case it is at $M_h \approx 10^{12}\,M_\odot$. There is also slight disagreement of all abundance matching results with the results of \citet{Reyes2012} (large green circles), who used a stacked weak lensing analysis of over a hundred thousand disk galaxies in SDSS separated in three bins of stellar mass. Regardless of the method used, however, there is a clear consensus that $M_\ast/M_h$ is much lower than the universal baryon fraction $f_b \approx 0.16$ (horizontal black dash-dotted line in figure~\ref{fig:am1}), at all halo masses. 


\begin{figure}[htbp]
\includegraphics[scale=0.7]{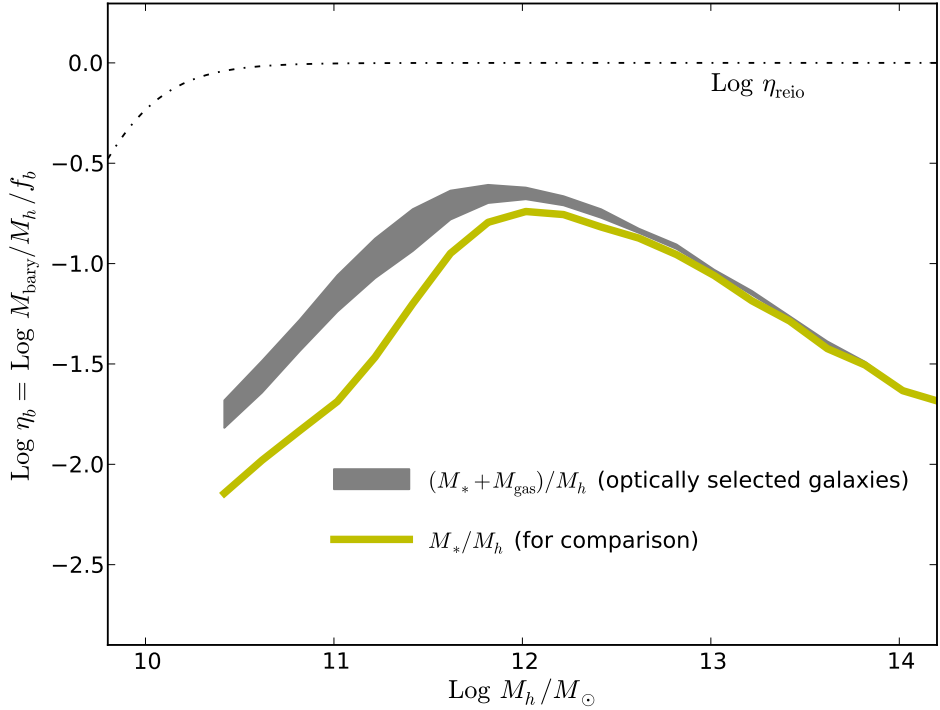}
\caption{The baryon fraction of galaxies, including stars and atomic gas, as a function of their host halo mass ($M_b / M_h - M_h$ relation). The gray shaded area shows the results of an abundance matching analysis of our optically-selected sample. Its boundaries correspond to two extreme assumptions for the gas content of galaxies detected optically but not in HI: i) galaxies that are not detected in HI contain no gas (lower boundary) and ii) galaxies that are not detected in HI contain the largest amount of gas that could have escaped detection from ALFALFA (upper boundary). The thick solid yellow line is the $M_\ast / M_h$-$M_h$ relation (same as in figure~\ref{fig:am1}), and is shown for comparison. The dotted-dashed line shows the baryon fraction that \citet{Okamoto2008} predict  based on hydrodynamic simulations that include cosmic reionisation.     }
\label{fig:am2}
\end{figure}

Let us now examine the results for the $M_b/M_h$ - $M_h$ relation (Fig. \ref{fig:am2}). The gray shaded area shows the relation derived from our optically-selected sample, matched to a halo mass function that includes both haloes and subhaloes: its upper and lower envelopes correspond to the distribution for $M_b^{max}$ and $M_b^{min}$, respectively, as defined in \S~\ref{sec:derived}. The thick gold solid line represents the relation for the stellar mass (same as in Figure~\ref{fig:am1}) and has been added for reference. Figure~\ref{fig:am2} puts in evidence the fact that $\eta_b$ decreases monotonically with decreasing halo mass, despite the fact that atomic gas contributes progressively more to the baryonic mass budget of galaxies.


\begin{figure}[htbp]
\includegraphics[scale=0.7]{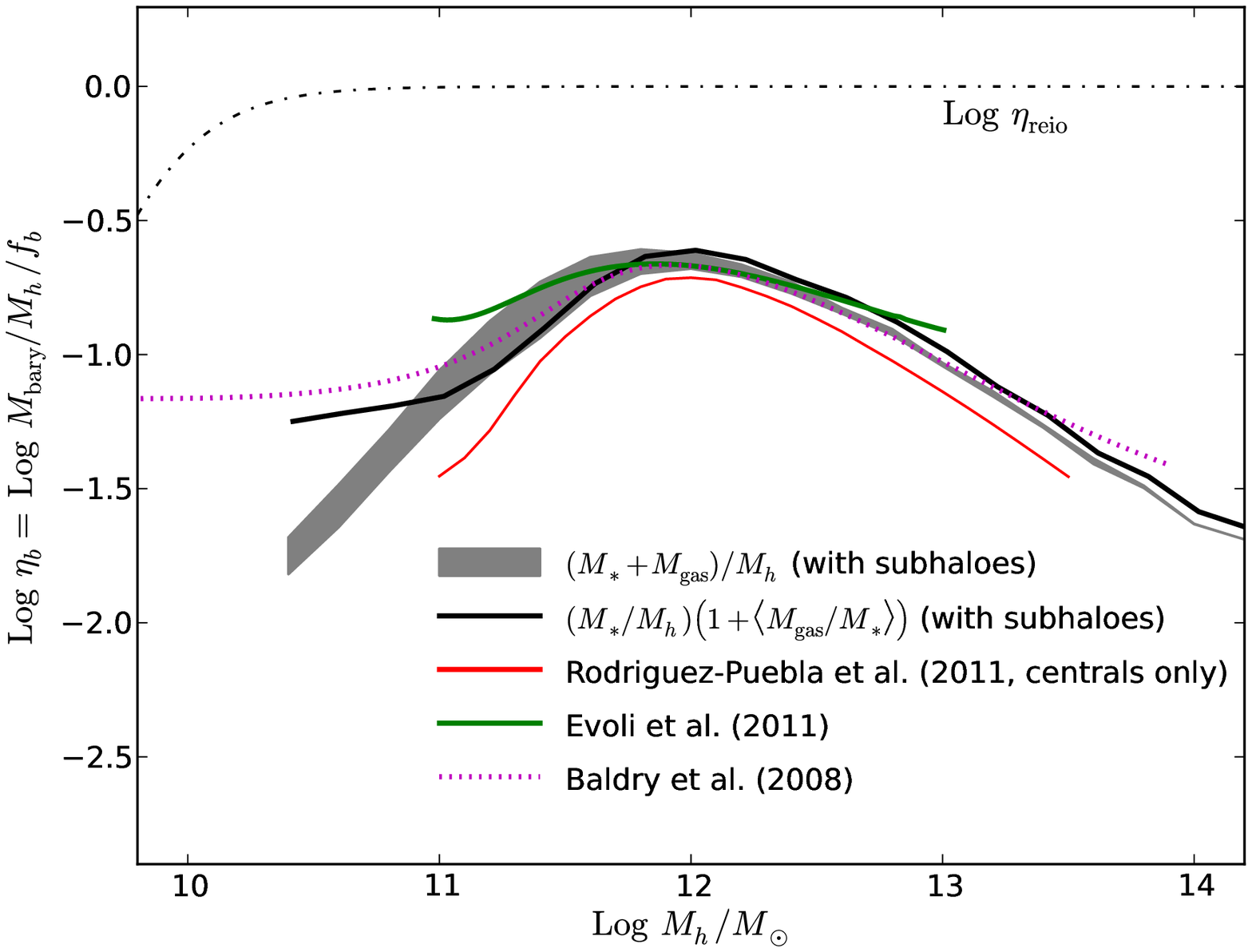}
\caption{The gray shaded area represents the $M_b / M_h$ - $M_h$ relation derived in this work from our optically-selected sample (same as Fig. \ref{fig:am2}). The magenta dotted line and the green solid lines show the $M_b/M_h$ - $M_h$ relations that \citet{Baldry2008} and \citet{Evoli2011} derived from their $M_\ast/M_h$ - $M_h$ relations (Fig.~\ref{fig:am1}), using the mean gas-to-stellar mass ratio as a function of stellar mass to account for the gas content of galaxies. The black solid line correspond to the results obtained from our optically-selected sample when we use the same method. The red solid line represents the result of \citet{Rodriguez2011}, who used separate $f_{HI}$ - $M_\ast$ relations for blue and red central galaxies. Lastly, the thin black dotted-dashed line is the same as Fig. \ref{fig:am3} and shows the baryon fraction that \citet{Okamoto2008} predict  based on hydrodynamic simulations that include cosmic reionisation.}
\label{fig:am3}
\end{figure}

In Fig. \ref{fig:am3} we compare our results to those by \citet{Baldry2008} and \citet{Evoli2011}, who derived their  $M_b/M_h$ - $M_h$ relations from the equation

\begin{equation}
{M_b\over M_h}={M_\ast \over M_h}\left(1+\langle \frac{M_{gas}}{M_\ast} \rangle \right).
\label{eqn:MbMh}
\end{equation}

\noindent
In Eqn.~\ref{eqn:MbMh} the baryonic mass is computed from the stellar mass, using the mean gas-to-stellar mass ratio for galaxies as a function of stellar mass. To enable a cleaner comparison, we have made the exercise of re-deriving the $M_b/M_h$ - $M_h$ relation from our $M_\ast/M_h$ - $M_h$ relation, using the same procedure followed by \citet{Baldry2008} (as in Eq.~\ref{eqn:MbMh}). The result is shown by the black solid line in Fig.~\ref{fig:am3}. The main difference between the results obtained by using individual galaxy gas masses (gray shaded region) and by adopting a mean gas-to-stellar mass ratio (black solid line) seems to be an artificial flattening of the $M_b / M_h$ relation at low masses. This is probably due to the fact that the latter method ignores the large scatter of galactic $M_{HI} / M_\ast$ values from the mean, and leads to the incorrect interpretation that the baryon retention fraction ($\eta_b$) of low-mass halos asymptotes to some fixed value. The red solid line in Fig.~\ref{fig:am3} corresponds to the result of \citet{Rodriguez2011}, who used a separate mean $M_{HI} / M_\ast$ - $M_\ast$ relation for blue and red galaxies. Their result\footnotemark{} shows no signs of flattening at low masses, however the relation only extends down to $M_h = 10^{11} \, M_\odot$. Independently of the used method however, all results point to values of $\eta_b$ that are well below unity, and cannot be explained by the effects of cosmic reionization alone (black dotted-dashed line in Fig.~\ref{fig:am3}; see Sec.~\ref{sec:discussion} for details). 

\footnotetext{Note that the \citet{Rodriguez2011} result refers to central galaxies only.}

\begin{figure}[htbp]
\includegraphics[scale=0.7]{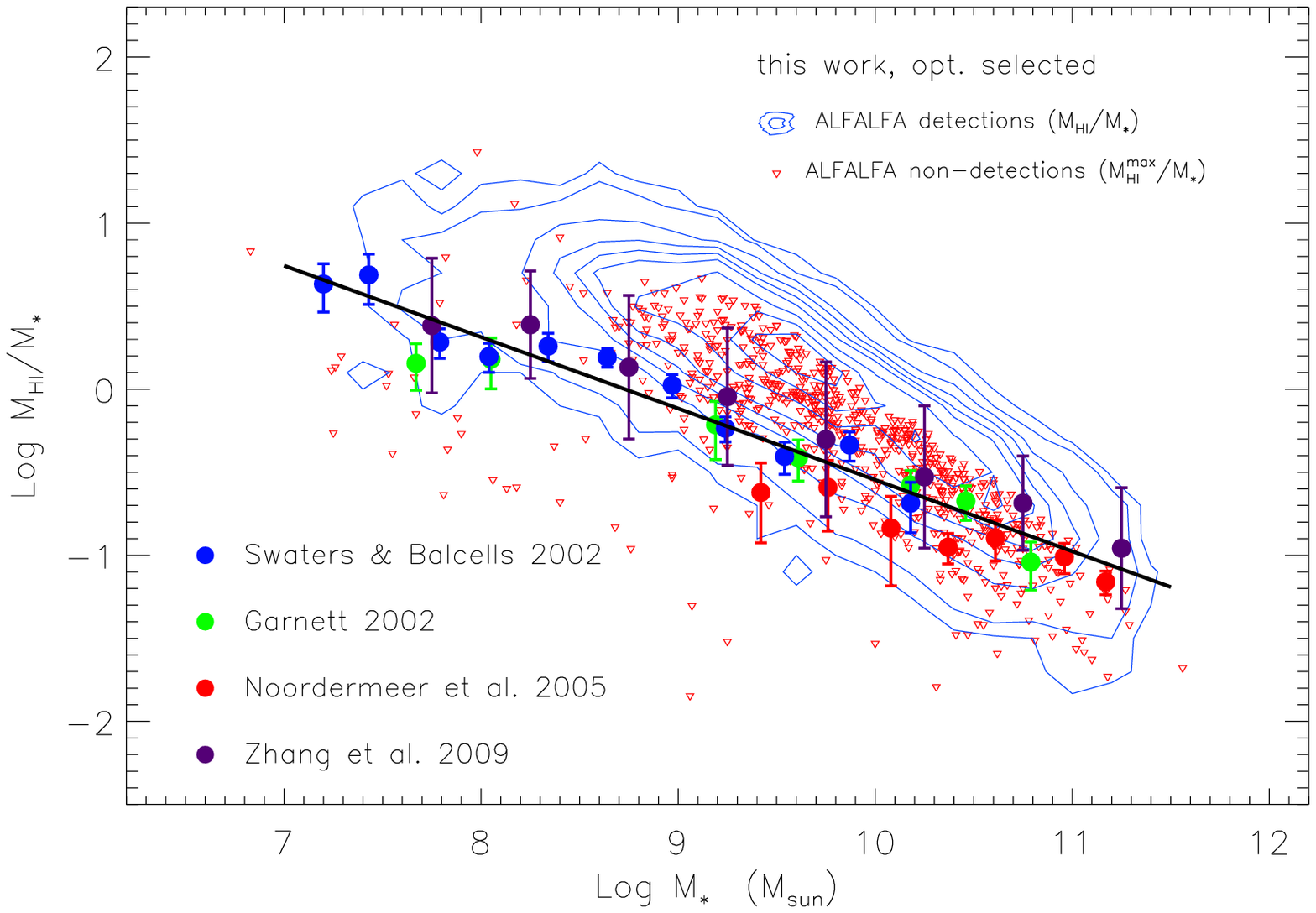}
\caption{HI-to-stellar mass ratio vs. galaxy stellar mass. The symbols with error bars show the  
results of \citet[blue circles]{SB2002}, \citet[green circles]{Garnett2002}, \citet[red circles]{Noordermeer2005} and \citet[magenta circles]{Zhang2009} for the average HI-to-stellar mass ratio in bins of stellar mass. The black line is a power-law fit to these data points. The blue contours represent the distribution of HI-fraction ($M_{HI} / M_*$) for the galaxies in our optically-selected sample that are detected by ALFALFA, while the red inverted triangles are maximum HI-fractions ($M_{HI}^{max} / M_\ast$) for a representative subset of our optically-selected galaxies that are not detected by ALFALFA. }
\label{fig:gsmr}
\end{figure}

The $\langle M_{gas} / M_\ast \rangle$ relation that we use to evaluate the right hand side of Eqn.~\ref{eqn:MbMh} (black solid line in Fig.~\ref{fig:gsmr}) is a power-law fit to the $M_{HI}/M_\ast$ data by \citet{SB2002}, \citet{Garnett2002}, \citet{Noordermeer2005} and \citet{Zhang2009}. The best fit relation, plotted as a thin solid black line in Fig.~\ref{fig:gsmr}, is given by $\log (M_{HI} / M_\ast) = -0.43 \, \log(M_\ast/M_\odot) + 3.75$. The blue contours in Fig.~\ref{fig:gsmr} represent the distribution of $M_{HI}/M_\ast$ values for the galaxies in our optically-selected sample that are detected by ALFALFA. Since the ALFALFA survey is a blind HI survey with a fixed integration time per pointing, the ALFALFA distribution is expected to lie above the data obtained by pointed observations of optically-selected galaxies (as in \citealp{SB2002, Garnett2002, Noordermeer2005}). The inverted red triangles correspond to the maximum HI-fraction ($M_{HI}^{max} / M_\ast$) for a representative subsample of our optically-selected galaxies that lack an ALFALFA detection. Note that these upper limits are also systematically higher than the relationship indicated by the power-law fit. 




\subsection{Discussion    \label{sec:discussion}}

The main result of this paper is the large ``gap'' between the present-day baryon fraction of galaxies in low-mass halos and the cosmic value ($f_b \approx 0.16$), which is present even when the atomic gas content of the galaxies is taken into account. This result is not contrived, given that atomic gas dominates the baryonic mass budget of galaxies with $M_h \lesssim 10^{11} \, M_\odot$. Moreover, the low-mass behavior is in disagreement with previous studies \citep[e.g.][]{Baldry2008, Evoli2011}, who find that the $\eta_b$ - $M_h$ relation flattens out at low masses, and approaches a roughly constant value of $\eta_b \approx 10\%$. These previous results would then require an exceptionally low efficiency of gas-to-stars conversion in low mass systems to  explain the observed values of $\eta_\ast$, which decrease monotonically.

In Fig.~\ref{fig:am2} we compare our result for the baryon fraction of halos to the predictions at $z=0$ of a cosmological hydrodynamic simulation that includes heating from a photoionizing UV background \citep[black dotted-dashed curve]{Okamoto2008}. At the high-mass end ($M_h \gtrsim 10^{13} \, M_\odot$), the low values of $\eta_b$ can be explained by the fact that infalling gas is shock-heated \citep{Keres2005, DB2006}, and is kept hot by feedback from active galactic nuclei \citep{Croton2006, Cattaneo2006, Bower2006}. This picture is supported by considerable observational evidence in the case of X-ray groups and clusters, but the situation within individual galaxies is not so well understood (see \citealp{Cattaneo2009} for a review).

At the low mass end, photoionization heating is expected to become important, since the intergalactic medium is too hot to fall into the shallow potential wells of haloes with $M_h < 10^{10} \, M_\odot$, and their baryon fraction is heavily suppressed. However, this process alone cannot account for the shape of the $\eta_b$ - $M_h$ relation at low masses, especially for the onset of a sharp decrease in the $M_b/M_h$ ratio at relatively large halo masses ($M_h \approx 10^{11.5} \, M_\odot$). Additional feedback is therefore needed, usually attributed to the ejection of baryons by stellar-driven winds. Semianalytic models of galaxy formation based on this assumption reproduce a good fit to luminosity functions in the local universe  \citep{Guo2011, BB2010, Somerville2008, Cattaneo2006} but the implied outflow rates are enormous. To reproduce the result presented in Figure~\ref{fig:am2}, the outflow rate in a halo with $M_h \approx 10^{10.3} \,M_\odot$  must be of order a hundred times higher than the star formation rate. It is difficult at present to reproduce such outflow rates in hydrodynamic simulations. Moreover, observational estimates place the total mass of outflows in normal star-forming galaxies at approximately the same level as the galaxies' final stellar mass \citep{Zahid2012}. Even in the most extreme observational cases, the ``mass-loading factor'' (the ratio of mass loss rate due to outflows over the star formation rate) is estimated to be $\lesssim 10$ \citep{Newman2012}. Therefore, explaining in detail the mechanisms responsible for the very low $\eta_b$ values found in low-mass galaxies seems to be a fundamental challenge for models of galaxy formation in a $\Lambda$CDM cosmological context.

Recent cosmological hydrodynamic simulations have challenged this statement, by managing to reproduce ``realistic'' galaxies whose properties satisfy a number of observational constraints. For example, the high-resolution Eris simulation \citep{Guedes2011} has managed to produce a Milky-Way type object with values of $\eta_\ast$ in agreement with those that we see in Fig.~\ref{fig:am2}. More recently, \citet{McCarthy2012} has managed to reproduce a population of $\sim$1000 simulated galaxies with low stellar-to-halo mass ratios ($\eta_\ast \lesssim 0.05$ at $M_h \approx 10^{11.3} \, M_\odot$), in accordance to observations. Notice, however, that this simulation uses a kinetic rather than thermal wind model, in which the initial wind speed is $600 \, {\rm km/s}$ and the initial mass-loading factor is a factor of four, by construction. Lastly, the work of \citet{Guedes2011} has been extended to lower masses by \citet{Brook2012}, who managed to produce a pair of dwarf galaxies ($M_h \approx 10^{10.8} \, M_\odot$), which obey the observed ``baryonic Tully-Fisher'' relation and are therefore expected to have the correct baryon-to-halo mass fractions.

While these studies indicate that we may be heading toward a solution of the discrepancy between the observed and the expected baryon content of dark matter halos, many questions (e.g. the expected outflow rates and re-accretion timescales) remain unanswered. Therefore, explaining in detail the mechanisms responsible for the very low $\eta_b$ values found in low-mass galaxies remains an open problem for studies of galaxy formation in a $\Lambda$CDM Universe. Furthermore, our measurement of $M_b/M_\ast$ provides an additional constraint with which cosmological hydrodynamical simulations and semianalytic models will have to confront.

\section{Conclusions     \label{sec:conclusion}}

We use optical data from the seventh data release of the Sloan Digital Sky Survey (SDSS DR7) and 21cm emission-line data from the Arecibo Legacy Fast ALFA (ALFALFA) survey to measure the ``baryonic mass'' (defined as $M_b = M_\ast + 1.4 \, M_{HI}$) of galaxies in the local universe, and determine the $z=0$ baryon mass function (BMF). We use both an HI-selected and an optically-selected sample (7195 and 22587 galaxies respectively) drawn from the same volume, in order to address the effects of sample selection on the derived distributions. We find that the main difference consists of the optically-selected stellar mass function (SMF) being systematically larger at high-masses than the HI-selected SMF, and find that this difference carries over to the high-mass end of the BMF (see Fig. \ref{fig:opt_vs_hi_gsmf} \& \ref{fig:opt_vs_hi_bmf}).

We combine the obtained mass distributions with the halo mass function in a WMAP3 $\Lambda$CDM cosmology, to obtain average values of $M_\ast / M_h$ and $M_b / M_h$ as a function of halo mass (Fig. \ref{fig:am1} \& \ref{fig:am2}). Our most important result is that low-mass halos seem to have very low galactic baryon fractions compared to the cosmic value ($f_b = \Omega_b / \Omega_m \approx 0.16$), even when their atomic gas content is taken into account; for example, the average baryon fraction of halos with $M_h = 10^{10.3} \, M_\odot$ is just 2\% of the cosmic value ($\eta_b \approx 0.02$), and displays a monotonically decreasing trend. This result contrasts with previous indirect measurements of the BMF \citep{Baldry2008, Evoli2011}, which pointed to an approximately constant value of $\eta_b \approx 0.10$ at the low halo-mass end.

Such very low values of $\eta_b$ are difficult to reconcile with current models of galaxy formation. Photoionization heating in the early universe suppresses the baryonic content of halos only at $M_h \lesssim 10^{10} \, M_\odot$ \citep{Okamoto2008}, but this mass is more than an order of magnitude smaller than what is required by our result. Therefore, additional feedback mechanisms, such as baryon blowout by supernova explosions, must be present and must be extremely efficient. It is not yet clear whether hydrodynamic simulations or observational results can accommodate such intense galactic outflows in low mass halos. As a result, the observed $\eta_b$ - $M_h$ relation remains difficult to explain, and may represent a challenge to our understanding of galaxy formation and/or the properties of dark matter.

\acknowledgements
\noindent
\textbf{Acknowledgements}

\noindent
The authors acknowledge the work of the entire ALFALFA collaboration team in observing, flagging, and extracting the catalog of galaxies used in this work. The ALFALFA team at Cornell is supported by NSF grants AST-0607007 and AST-1107390 to RG and MPH and by grants from the Brinson Foundation. EP would also like to thank the Onassis Foundation for partial support. AC thanks Alexie Leauthaud for providing many of the data points used to make Fig.~\ref{fig:am1}.

\noindent
Funding for the SDSS and SDSS-II has been provided by the Alfred P. Sloan Foundation, the participating institutions, the National Science Foundation, the US Department of Energy, the NASA, the Japanese Monbukagakusho, the Max Planck Society and the Higher Education Funding Council for England. The SDSS Web Site is \texttt{http://www.sdss.org}. The SDSS is managed by the Astrophysical Research Consortium for the participating institutions. 


\begin{thebibliography}



\bibitem[Reddick et al.(2012)]{Reddick2012} Reddick, R.~M., 
Wechsler, R.~H., Tinker, J.~L., \& Behroozi, P.~S.\ 2012, arXiv:1207.2160 

\bibitem[Brook et al.(2012)]{Brook2012} Brook, C.~B., Stinson, 
G., Gibson, B.~K., Wadsley, J., \& Quinn, T.\ 2012, \mnras, 424, 1275 

\bibitem[Rodr{\'{\i}}guez-Puebla et al.(2012)]{Rodriguez2012} 
Rodr{\'{\i}}guez-Puebla, A., Drory, N., 
\& Avila-Reese, V.\ 2012, \apj, 756, 2 

\bibitem[Zahid et al.(2012)]{Zahid2012} Zahid, H.~J., Dima, 
G.~I., Kewley, L.~J., Erb, D.~K., \& Dave, R.\ 2012, arXiv:1207.5509 

\bibitem[Hallenbeck et al.(2012)]{Hallenbeck2012} Hallenbeck, G., 
Papastergis, E., Huang, S., et al.\ 2012, \aj, 144, 87 

\bibitem[Huang et al.(2012)]{Huang2012b} Huang, S., Haynes, M.~P., 
Giovanelli, R., \& Brinchmann, J.\ 2012, arXiv:1207.0523 
 
\bibitem[McCarthy et al.(2012)]{McCarthy2012} McCarthy, I.~G., 
Schaye, J., Font, A.~S., et al.\ 2012, arXiv:1204.5195 

\bibitem[Wetzel et al.(2012)]{Wetzel2012} Wetzel, A.~R., Tinker, 
J.~L., \& Conroy, C.\ 2012, \mnras, 3128 

\bibitem[Martin et al.(2012)]{Martin2012} Martin, A.~M., 
Giovanelli, R., Haynes, M.~P., \& Guzzo, L.\ 2012, \apj, 750, 38 

\bibitem[Reyes et al.(2012)]{Reyes2012} Reyes, R., Mandelbaum, R., Gunn, J.~E., Nakajima, R., Seljak, U., Hirata, C.~M., 2012, \mnras

\bibitem[Pforr et al.(2012)]{Maraston2012} Pforr, J., Maraston, C., 
\& Tonini, C.\ 2012, arXiv:1203.3548 

\bibitem[Baldry et al.(2012)]{Baldry2012} Baldry, I.~K., Driver, 
S.~P., Loveday, J., et al.\ 2012, \mnras, 421, 621 

\bibitem[Newman et al.(2012)]{Newman2012} Newman, S.~F., Shapiro 
Griffin, K., Genzel, R., et al.\ 2012, \apj, 752, 111 

\bibitem[Huang et al.(2012)]{Huang2012a} Huang, S., Haynes, M.~P., 
Giovanelli, R., et al.\ 2012, \aj, 143, 133 

\bibitem[Leauthaud et al.(2012)]{Leauthaud2012} Leauthaud, A., 
Tinker, J., Bundy, K., et al.\ 2012, \apj, 744, 159 

\bibitem[McGaugh(2012)]{McGaugh2012} McGaugh, S.~S.\ 2012, \aj, 
143, 40 

\bibitem[Hall et al.(2012)]{Hall2012} Hall, M., Courteau, S., 
Dutton, A.~A., McDonald, M., \& Zhu, Y.\ 2012, \mnras, 3501 




\bibitem[Rodr{\'{\i}}guez-Puebla et al.(2011)]{Rodriguez2011} 
Rodr{\'{\i}}guez-Puebla, A., Avila-Reese, V., Firmani, C., 
\& Col{\'{\i}}n, P.\ 2011, \rmxaa, 47, 235 

\bibitem[Taylor et al.(2011)]{Taylor2011} Taylor, E.~N., Hopkins, 
A.~M., Baldry, I.~K., et al.\ 2011, \mnras, 418, 1587 

\bibitem[Stinson et al.(2011)]{Stinson2011} Stinson, G., Brook, C., 
Prochaska, J.~X., et al.\ 2011, arXiv:1112.1698 

\bibitem[Guedes et al.(2011)]{Guedes2011} Guedes, J., Callegari, 
S., Madau, P., \& Mayer, L.\ 2011, \apj, 742, 76 

\bibitem[Evoli et al.(2011)]{Evoli2011} Evoli, C., Salucci, P., Lapi, A., Danese, L.., 2011, \apj, 743, 45

\bibitem[More et al.(2011)]{More2011} More, S., van den Bosch, F.~C., Cacciato, M., Skibba, R., Mo, H.~J., Yang, X., 2011, \mnras, 410, 210

\bibitem[Guo et al.(2011)]{Guo2011} Guo, Q., et al., 2011, \mnras, 413, 101

\bibitem[Haynes et al.(2011)]{Haynes2011} Haynes, M.~P., 
Giovanelli, R., Martin, A.~M., et al.\ 2011, \aj, 142, 170 

\bibitem[Papastergis et al.(2011)]{Papastergis2011} Papastergis, E., 
Martin, A.~M., Giovanelli, R., \& Haynes, M.~P.\ 2011, \apj, 739, 38


\bibitem[Saintonge et al.(2011)]{Saintonge2011} Saintonge, A., 
Kauffmann, G., Kramer, C., et al.\ 2011, \mnras, 415, 32 

\bibitem[Komatsu et al.(2011)]{Komatsu2011} Komatsu, E., Smith, 
K.~M., Dunkley, J., et al.\ 2011, \apjs, 192, 18 



\bibitem[Driver 
\& Robotham(2010)]{DR2010} Driver, S.~P., \& Robotham, A.~S.~G.\ 2010, \mnras, 407, 2131 

\bibitem[Leauthaud et al.(2010)]{Leauthaud2010} Leauthaud, A., et al., 2010, \apj, 709, 97

\bibitem[Martin et al.(2010)]{Martin2010} Martin, A.~M., 
Papastergis, E., Giovanelli, R., et al.\ 2010, \apj, 723, 1359 

\bibitem[Dutton et al.(2010)]{Dutton2010} Dutton, A.~A., Conroy, 
C., van den Bosch, F.~C., Prada, F., \& More, S.\ 2010, \mnras, 407, 2 

\bibitem[Benson \& Bower(2010)]{BB2010} Benson, A.~J., Bower, R., 2010, \mnras, 405, 1573

\bibitem[Behroozi et al.(2010)]{Behroozi2010} Behroozi, P.~S., 
Conroy, C., \& Wechsler, R.~H.\ 2010, \apj, 717, 379 

\bibitem[Guo et al.(2010)]{Guo2010} Guo, Q., White, S., Li, C., 
\& Boylan-Kolchin, M.\ 2010, \mnras, 404, 1111 

\bibitem[Catinella et al.(2010)]{Catinella2010} Catinella, B., 
Schiminovich, D., Kauffmann, G., et al.\ 2010, \mnras, 403, 683 

\bibitem[Moster et al.(2010)]{Moster2010} Moster, B.~P., 
Somerville, R.~S., Maulbetsch, C., et al.\ 2010, \apj, 710, 903 



\bibitem[West et al.(2009)]{West2009} West, A.~A., 
Garcia-Appadoo, D.~A., Dalcanton, J.~J., et al.\ 2009, \aj, 138, 796 


\bibitem[Tweed et al.(2009)]{Tweed2009} Tweed, D., Devriendt, J., Blaizot, J., Colombi, S., \& Slyz, A.\ 2009, \aap, 506, 647 

\bibitem[Abazajian et al.(2009)]{Abazajian2009} Abazajian, K.~N., 
Adelman-McCarthy, J.~K., Ag{\"u}eros, M.~A., et al.\ 2009, \apjs, 182, 543 

\bibitem[Li \& White(2009)]{LW2009} Li, C., \& White, S.~D.~M.\ 2009, \mnras, 398, 2177 

\bibitem[Leroy et al.(2009)]{Leroy2009} Leroy, A.~K., Walter, F., 
Bigiel, F., et al.\ 2009, \aj, 137, 4670 

\bibitem[Zhang et al.(2009)]{Zhang2009} Zhang, W., Li, C., Kauffmann, G., Zou, H., Catinella, B., Shen, S., Guo, Q., Chang, R., 2009, \mnras, 397, 1243

\bibitem[Yang et al.(2009)]{Yang2009} Yang, X., Mo, H.~J., 
\& van den Bosch, F.~C.\ 2009, \apj, 695, 900 

\bibitem[Cattaneo et al.(2009)]{Cattaneo2009} Cattaneo, A., Faber, 
S.~M., Binney, J., et al.\ 2009, \nat, 460, 213 


 
\bibitem[Somerville et al.(2008)]{Somerville2008} Somerville, R.~S., Hopkins, P.~F., Cox, T.~J., Robertson, B.~E., Hernquist, L., 2008, \mnras, 481, 506

\bibitem[Okamoto et al.(2008)]{Okamoto2008} Okamoto, T., Gao, L., \& Theuns, T.\ 2008, \mnras, 390, 920 

\bibitem[Baldry et al.(2008)]{Baldry2008} Baldry, I.~K., 
Glazebrook, K., \& Driver, S.~P.\ 2008, \mnras, 388, 945 



\bibitem[Cattaneo et al.(2007)]{Cattaneo2007} Cattaneo, A., Blaizot, 
J., Weinberg, D.~H., et al.\ 2007, \mnras, 377, 63 

\bibitem[Spergel et al.(2007)]{Spergel2007} Spergel, D.~N., Bean, 
R., Dor{\'e}, O., et al.\ 2007, \apjs, 170, 377 

\bibitem[Conroy et al.(2007)]{Conroy2007} Conroy, C., et al., 2007, \apj, 654, 153

\bibitem[Salim et al.(2007)]{Salim2007} Salim, S., Rich, R.~M., 
Charlot, S., et al.\ 2007, \apjs, 173, 267 

\bibitem[Springob et al.(2007)]{Springob2007} Springob, C.~M., Masters, K.~L., Haynes, M.~P., Giovanelli, R., Marinoni, C., 2007, \apjs, 172, 599

\bibitem[Pizagno et al.(2007)]{Pizagno2007} Pizagno, J., 2007, \aj, 134, 945

\bibitem[Hoekstra(2007)]{Hoekstra2007} Hoekstra, H., 2007, \mnras, 379, 317

\bibitem[Panter et al.(2007)]{Panter2007} Panter, B., Jimenez, R., 
Heavens, A.~F., \& Charlot, S.\ 2007, \mnras, 378, 1550 




\bibitem[Weinmann et al.(2006)]{Weinmann2006} Weinmann, S.~M., van 
den Bosch, F.~C., Yang, X., et al.\ 2006, \mnras, 372, 1161 

\bibitem[Cattaneo et al. (2006)]{Cattaneo2006} Cattaneo, A., Dekel, A., Devriendt, J., Guiderdoni, B., Blaizot, J., 2006, \mnras, 370, 1651

\bibitem[Geha et al.(2006)]{Geha2006} Geha, M., Blanton, M.~R., Masjedi, M., West, A.~A., 2006, \apj, 653, 240

\bibitem[Bower et al.(2006)]{Bower2006} Bower, R.~G., Benson, A.~J., Malbon, R., et al.\ 2006, \mnras, 370, 645 

\bibitem[Croton et al.(2006)]{Croton2006} Croton, D.~J., Springel, V., White, S.~D.~M., et al.\ 2006, \mnras, 365, 11 

\bibitem[Mandelbaum et al.(2006)]{Mandelbaum2006} Mandelbaum, R., Seljak, U., Kauffmann, G., Hirata, C.~M., Brinkmann, J., 2006, \mnras, 368, 715

\bibitem[Dekel \& Birnboim(2006)]{DB2006} Dekel, A., \& Birnboim, Y.\ 2006, \mnras, 368, 2 

\bibitem[Shankar et al. (2006)]{Shankar2006} Shankar, F., Lapi, A., Salucci, P., De Zotti, G., Danese, L. \ 2006, \apj, 643, 14



\bibitem[Springel(2005)]{Springel2005} Springel, V.\ 2005, \mnras, 
364, 1105 

\bibitem[Kere{\v s} et al.(2005)]{Keres2005} Kere{\v s}, D., 
Katz, N., Weinberg, D.~H., \& Dav{\'e}, R.\ 2005, \mnras, 363, 2 

\bibitem[Gallazzi et al.(2005)]{Gallazzi2005} Gallazzi, A., Charlot, S., Brinchmann, J., White, S.~D.~M., \& Tremonti, C.~A.\ 2005, \mnras, 362, 41 

\bibitem[Blanton et al.(2005)]{Blanton2005} Blanton, M.~R., 
Schlegel, D.~J., Strauss, M.~A., et al.\ 2005, \aj, 129, 2562 

\bibitem[Zwaan et al.(2005)]{Zwaan2005} Zwaan, M.~A., Meyer, 
M.~J., Staveley-Smith, L., \& Webster, R.~L.\ 2005, \mnras, 359, L30 

\bibitem[Noordermeer et 
al.(2005)]{Noordermeer2005} Noordermeer, E., van der Hulst, J.~M., Sancisi, R., Swaters, R.~A., \& van Albada, T.~S.\ 2005, \aap, 442, 137 

\bibitem[Masters(2005)]{Masters2005} Masters, K.~L.\ 2005, 
Ph.D.~Thesis  



\bibitem[Somerville et al.(2004)]{Somerville2004} Somerville, R.~S., 
Lee, K., Ferguson, H.~C., et al.\ 2004, \apjl, 600, L171 

\bibitem[Aubert et al.(2004)]{Aubert2004} Aubert, D., Pichon, C., 
\& Colombi, S.\ 2004, \mnras, 352, 376 

\bibitem[Reynolds(2004)]{Reynolds2004} Reynolds, R.~J.\ 2004, 
Advances in Space Research, 34, 27 

\bibitem[Brinchmann et al.(2004)]{Brinchmann2004} Brinchmann, J., 
Charlot, S., White, S.~D.~M., et al.\ 2004, \mnras, 351, 1151 

\bibitem[Glazebrook et al.(2004)]{Glazebrook2004} Glazebrook, K., 
Abraham, R.~G., McCarthy, P.~J., et al.\ 2004, \nat, 430, 181 

\bibitem[Masters et al.(2004)]{Masters2004} Masters, K.~L., Haynes, 
M.~P., \& Giovanelli, R.\ 2004, \apjl, 607, L115 

\bibitem[Vale \& Ostriker(2004)]{ValeOstriker2003} Vale, A., Ostriker, J.~P., \ 2004, \mnras, 353, 189


\bibitem[Bell et al.(2003)]{Bell2003} Bell, E.~F., McIntosh, 
D.~H., Katz, N., \& Weinberg, M.~D.\ 2003, \apjs, 149, 289 

\bibitem[Chabrier(2003)]{Chabrier2003} Chabrier, G.\ 2003, \pasp, 
115, 763

\bibitem[Bruzual 
\& Charlot(2003)]{BC2003} Bruzual, G., \& Charlot, S.\ 2003, \mnras, 344, 1000 

\bibitem[Zwaan et al.(2003)]{Zwaan2003} Zwaan, M.~A., 
Staveley-Smith, L., Koribalski, B.~S., et al.\ 2003, \aj, 125, 2842 

\bibitem[Kauffmann et al.(2003)]{Kauffmann2003} Kauffmann, G., 
Heckman, T.~M., White, S.~D.~M., et al.\ 2003, \mnras, 341, 33 



\bibitem[Boselli et 
al.(2002)]{Boselli2002} Boselli, A., Lequeux, J., \& Gavazzi, G.\ 2002, \aap, 384, 33 

\bibitem[Swaters 
\& Balcells(2002)]{SB2002} Swaters, R.~A., \& Balcells, M.\ 2002, \aap, 390, 863 

\bibitem[Garnett(2002)]{Garnett2002} Garnett, D.~R.\ 2002, \apj, 
581, 1019 

\bibitem[Marinoni \& Hudson(2002)]{MarinoniHudson2002} Marinoni, C., Hudson, M.~J., \ 2002, \apj, 569, 101



\bibitem[Cole et al.(2001)]{Cole2001} Cole, S., Norberg, P., 
Baugh, C.~M., et al.\ 2001, \mnras, 326, 255 



\bibitem[Salucci \& Persic(1999)]{SP1999} Salucci, P., \& Persic, M.\ 1999, \mnras, 309, 923 


 



\bibitem[Efstathiou et al.(1988)]{Efstathiou1988} Efstathiou, G., 
Ellis, R.~S., \& Peterson, B.~A.\ 1988, \mnras, 232, 431 



\bibitem[Schmidt(1968)]{Schmidt1968} Schmidt, M.\ 1968, \apj, 151, 
393 



\end{thebibliography}
\end{document}